\documentclass[draft]{agujournal2019}
\usepackage{lmodern}
\usepackage{url} 
\usepackage{lineno}
\usepackage{soul}

\usepackage{amsmath} 
\usepackage{comment}
\usepackage[normalem]{ulem}
\newcommand{\kiab}[1]{\left \langle #1 \right \rangle}
\newcommand{\kirb}[1]{\left ( #1 \right )}
\newcommand{\kisb}[1]{\left [ #1 \right ]}
\newcommand{\kicb}[1]{\left \{ #1 \right \}}
\newcommand{\ud}{\,\mathrm{d}}
\newcommand{\kitens}[1]{\overline{\overline{#1}}}

\draftfalse

\journalname{Journal of Advances in Modeling Earth Systems (JAMES)}

\begin{document}

\title{Accurate Column Moist Static Energy Budget in Climate Models. Part 1: Conservation Equation Formulation, Methodology, and Primary Results Demonstrated Using GISS ModelE3}

\authors{Kuniaki Inoue\affil{1,2}, Maxwell Kelley\affil{2}, Ann M. Fridlind\affil{2}, Michela Biasutti\affil{3}, Gregory S. Elsaesser\affil{2,4}}

\affiliation{1}{Center for Climate System Research, Columbia University, New York, NY, USA}
\affiliation{2}{NASA Goddard Institute for Space Studies, New York, NY, USA}
\affiliation{3}{Lamont-Doherty Earth Observatory, Columbia University, Palisades, NY, USA}
\affiliation{4}{Department of Applied Physics and Mathematics, Columbia University, New York, NY, USA}

\correspondingauthor{Kuniaki Inoue}{kuni.inoue22@gmail.com}

\begin{keypoints}
\item Seven intertwined factors causing substantial column MSE budget residuals are identified in ModelE3. 
\item The ``process increment method'' is implemented for accurately calculating the column MSE budget terms, enabling precise MSE budget analysis. 
\item Errors from vertical interpolation can reverse the sign of vertical MSE advection, underscoring the necessity for accurate computations. 
\end{keypoints}

\begin{abstract}
Column‐integrated moist static energy (MSE) budgets underpin theories of tropical convection and circulation, yet in reanalyses and climate models the budget rarely closes; residuals routinely match the leading terms and mask physical insights. This study derives an MSE conservation law that is strictly consistent with GISS ModelE3 and elucidates why conventional diagnostics fail. Multiple intertwined factors---the breakdown of the product rule upon discretization, effects of mass-filtering, mismatched flux and advective forms, numerical noise in diagnosed vertical velocity, asynchronous model output timing, and postprocessing including vertical interpolation and temporal averaging—leave significant residuals in both annual means and daily variability, even when raw 30-min model output is used. Residuals are even larger over land and along coastlines. To tackle this obstacle, this study implements the ``process increment method,'' which accurately computes the column MSE flux divergence by calculating the change in column-integrated internal energy, geopotential energy, and latent heats before and after applying the dynamics scheme. Furthermore, the calculated column flux divergence is decomposed into horizontal and vertical advective components. The most crucial finding is that vertical interpolation into pressure coordinates can introduce errors substantial enough to reverse the sign of vertical MSE advection in the warm-pool regions. In ModelE3, native-grid values show MSE import via vertical circulations, while values after interpolation into pressure coordinates indicate export. This discrepancy may prompt a reevaluation of vertical advection as an exporting mechanism and underscores the importance of precise MSE budget calculations.

\end{abstract}

\section*{Plain Language Summary}
Moist static energy (MSE), the sum of total specific enthalpy and geopotential, is a key metric in atmospheric science. However, calculating its budget in climate models is challenging, often leading to significant errors and residuals that hinder accurate analysis. Our study incorporates inline diagnostic routines into the NASA GISS climate model to calculate the MSE budget precisely. We found that large residuals persist for reasons that cannot be eliminated with conventional offline methods. To address the issue, we use a technique called the ``process increment method'' for calculating the MSE budget. This method measures the change in combined heat and geopotential energy in the atmosphere before and after specific model processes related to energy transport. The most important finding is that the change in vertical coordinates for analysis can introduce large errors, enough to alter the understanding of how energy moves in certain regions. Accurate calculations show that energy is being imported into the tropics via vertical circulations in our models, while traditional methods suggest it is being exported. This discrepancy may require rethinking the conventional understanding of energy movement and underscores the need for precise energy calculations.

\newpage
\section{Introduction}
One of the most widely used metrics in meteorology and climate science is moist static energy (MSE)\footnote{Throughout this study we include the ice‐phase contribution—often termed frozen MSE—but, for brevity, we simply refer to it as MSE.}, which represents the sum of total specific enthalpy and gravitational potential energy. MSE is approximately conserved in air parcels undergoing moist adiabatic displacements \cite<e.g.,>{riehl_heat_1958, betts_further_1974}. Its conservation properties are instrumental for deducing in-cloud characteristics, such as cloud buoyancy and convective mass fluxes \cite<e.g.,>[]{yanai_determination_1973, singh_influence_2013, romps_mse_2015, romps_clausiusclapeyron_2016, masunaga_convective_2016, peters_formula_2020, peters_evaluating_2021, peters_analytic_2023}, and these properties are the basis of some convective parameterizations \cite<e.g.,>[]{arakawa_interaction_1974,moorthi_relaxed_1992,romps_stochastic_2016}.

Atmospheric circulation is fundamentally a manifestation of global energy transport, essential for maintaining regional energy balance in spite of strong gradients in solar energy input. Predominantly, this energy transport occurs in the form of MSE transport. As a result, the global MSE budget exerts a substantial influence on global circulation patterns. In light of this, numerous studies have focused on providing accurate estimates of the energy budget \cite<e.g.,>[]{riehl_heat_1958, trenberth_global_1994, trenberth_using_1997, trenberth_atmospheric_2001, trenberth_accuracy_2002, trenberth_covariability_2003, trenberth_seamless_2003, fasullo_annual_2008, mayer_toward_2017, mayer_consistency_2021}, especially since MSE transport significantly impacts the most prominent hydroclimatic features: the position and width of the inter-tropical convergence zone (ITCZ) \cite<e.g.,>[]{kang_response_2008, kang_tropical_2009, donohoe_relationship_2013, bischoff_energetic_2014, marshall_oceans_2014, schneider_migrations_2014, byrne_energetic_2016, biasutti_global_2018}, as well as the position and intensity of the mid-latitude storm tracks \cite<e.g.,>[]{barpanda_using_2017, shaw_moist_2018}.

The column-integrated MSE budget is particularly crucial in analyzing tropical convective variability across various time scales; pioneering work on using MSE to understand mean tropical convective patterns was conducted by \citeA{neelin_modeling_1987}, with subsequent work additionally using column MSE to understand various tropical convective phenomena. These include: the development and propagation of the Madden-Julian Oscillation (MJO) \cite<e.g.,>[]{maloney_moist_2009, hannah_role_2011, andersen_moist_2011, pritchard_causal_2013, kim_propagating_2014, Sobel_mse, Inoue_MSE_2015, arnold_global-scale_2015, wolding_objective_2015, yokoi_intraseasonal_2015, yasunaga_spacetime_2018, gonzalez_distinct_2019, benedict_investigating_rad} and convectively coupled equatorial waves \cite<e.g.,>[]{peters_structure_2006,  Inoue_MSE_2015, sumi_moist_2016, adames_moisture_2018, yasunaga_spacetime_2018, gonzalez_distinct_2019, mayta_westward-propagating_2022, nakamura_convective_2022}; the formation and development of tropical cyclones \cite<e.g.,>[]{mcbride_observational_1981, wing_moist_2019, dirkes_process-oriented_2023} and convective self-aggregation \cite<e.g.,>[]{bretherton_energy-balance_2005, muller_detailed_2012, wing_physical_2014, arnold_global-scale_2015, bretherton_convective_2015, holloway_sensitivity_2016, wing_self-aggregation_2016, wing_convective_2017}; the life-cycle of ubiquitous tropical convective variability \cite<e.g.,>[]{masunaga_mechanism_2014, inoue_gross_2015, inoue_gross_2017, inoue_evidence_2021}; the mechanisms of regional monsoons \cite<e.g.,>[]{chou_ocean-atmosphere-land_2001, chou_mechanisms_2001, chou_mechanisms_2003, mohanty_australian_2024, mohanty_australian_2024b}; and the impacts of global warming on tropical precipitation \cite<e.g.,>[]{neelin_tropical_2003, chou_mechanisms_2004, ahmed_process_2023} and on regional monsoons \cite<e.g.,>[]{hill_moist_2017, hill_robust_2018, hill_theories_2019, smyth_characterizing_2020}.

Despite its importance and widespread use, column-integrated MSE budget analysis is marred by a major limitation: budgets derived from reanalysis data or model output almost never close. In principle, the vertically integrated MSE tendency, its advective component, and the relevant source terms should sum to zero; in practice, substantial residuals remain. As we show later, these residuals are often comparable in magnitude to the main budget terms, thereby complicating efforts to evaluate and understand the individual contributions of each budget term \cite<e.g.,>[]{kim_propagating_2014, hill_theories_2019, ren_intercomparison_2021}. In typical analyses, these residuals are handled in one of three ways: they are disregarded, quantified independently as a measure of uncertainty, or merged into the dynamical flux divergence.

To address this limitation, we have implemented an inline calculation of the column MSE budget in the NASA Goddard Institute for Space Studies Earth System Model, ModelE version 3 (ModelE3), as part of our internal development effort. We employ a straightforward ``process increment method'' that diagnoses the column MSE flux divergence—together with an indirect estimate of vertical MSE advection—so that the budget closes exactly, with no residual. Although the complexities of MSE budget computation vary across models, the framework is deliberately designed for easy adaptation to a wide range of model architectures.

This paper is organized as follows. Section~\ref{sec:derivation} derives a column-integrated MSE budget that is fully consistent with ModelE3 physics. Section~\ref{sec:model} describes the model configuration pertinent to this study. Section~\ref{sec:residual} quantifies residuals obtained from postprocessed output, highlighting their significance and the necessity for a precise computation. Section~\ref{sec:cause} pinpoints the fundamental causes of these residuals. Section~\ref{sec:accuracy} introduces two key concepts—numerical consistency and physical consistency—that underpin our approach to accurate MSE-budget closure. Section~\ref{sec:method} first illustrates an ineffective strategy: computing the flux divergence directly from native-grid output at the highest resolution, which nevertheless leaves substantial residuals. It then presents our process increment method, which yields an exact budget closure and an indirect but accurate estimate of vertical MSE advection. Section~\ref{sec:results} demonstrates how the new method improves budget accuracy and argues that similar implementations in other models would broadly enable more reliable MSE-budget analyses. Finally, Section~\ref{sec:summary} offers concluding remarks.

\section{Derivations of Energy Budget Equations}\label{sec:derivation}
To ensure accurate computation of the column MSE budget using a model, it is crucial to derive the conservation law in a manner that aligns consistently with the underlying physics of the model. This section is dedicated to deriving the column MSE budget equation, comprehensively incorporating all relevant budget terms and clearly delineating every assumption used in ModelE3 for transparency and clarity. For derivations that incorporate approximations, readers may refer to \citeA{peixoto_physics_1992, trenberth_global_1994, randall_introduction_2015, adames-corraliza_accuracy_2023}. Note that for simplicity, all subsequent derivations are conducted in the vertical $z$-coordinate system, with any changes to vertical coordinates clearly specified. However, the choice of vertical coordinates does not influence the equations derived.

\subsection{Key Assumptions}
In ModelE3, two essential assumptions underpin the derivation of the column MSE budget equation. First, within the \emph{resolved-scale} physics, ModelE3 does not consider the effects of water substances (vapor, liquid, and solid) on air mass. This results in treating the dry air density as equivalent to the total air density. Consequently, the ideal gas law in the resolved-scale physics omits the virtual effect of water vapor, as shown in the equation:
\begin{linenomath*}
\begin{equation}\label{eq:ideal_gas}
p = \rho R_d T \;,
\end{equation}
\end{linenomath*}
where $p$ is the pressure, $\rho$ is the (total) air density, $R_d$ is the gas constant for dry air, and $T$ is the temperature. Furthermore, due to the lack of distinction between the dry air mass and the total air mass in the model, mixing ratios (per unit dry mass) and mass fractions (per unit total mass) are considered equivalent.

Water substances still influence cloud buoyancy, but those effects are handled entirely within the \emph{unresolved‐scale} convective parameterization.  At that scale, the virtual effect of water vapor is fully included in the buoyancy calculation.  Such modular treatment is common in climate models: different physics packages may incorporate moisture in different ways.

Second, ModelE3 treats specific heats and latent heats as constants, neglecting their variation with moisture content and temperature. All relevant constants are summarized in Table~\ref{t:1}.

For derivations that include the variability of specific heats and latent heats, readers are directed to \citeA{romps_mse_2015, marquet_comments_2016, peters_evaluating_2021} for a non-hydrostatic atmosphere, and to \citeA{mayer_toward_2017, kato_regional_2021} for a hydrostatic atmosphere.

\begin{table}[htb]
\begin{center}
\caption{List of constants}\label{t:1}
\begin{tabular}{cccc}
  \hline
  Symbol & Meaning & Value & Unit \\
  \hline
  $R_d$ & Gas constant for dry air & 287.04873 & J~kg$^{-1}$~K$^{-1}$\\
  $c_p$ & Specific heat at constant pressure & 1002.88098 & J~kg$^{-1}$~K$^{-1}$ \\
  $c_v$ & Specific heat at constant volume & $c_p-R_d$ & J~kg$^{-1}$~K$^{-1}$ \\
  $L_v$ & Latent heat of vaporization & 2.50$\times$10$^6$ & J~kg$^{-1}$\\
  $L_f$ & Latent heat of fusion & 3.34$\times$10$^5$ &  J~kg$^{-1}$ \\
  $L_s$ & Latent heat of sublimation & $L_v+L_f$ &  J~kg$^{-1}$ \\
  $g$ & Gravitational acceleration & 9.80665 & m~s$^{-2}$\\
  \hline
\end{tabular}
\end{center}
\end{table}

\subsection{Derivation of the Moist Static Energy Budget Equation}\label{subsec:mse_first}
In ModelE3, the potential temperature is a prognostic variable integrated within the model, and its governing equation is expressed as:
\begin{linenomath*}
\begin{equation}\label{eq:potential_temp}
c_p\Pi\frac{D\theta}{Dt}  = -\frac{1}{\rho}\nabla\cdot\kirb{\mathbf{R}+\mathbf{F}_t} +L_v\mathcal{C}+L_f\mathcal{F} + L_s\mathcal{D} + \delta\,,
\end{equation}
\end{linenomath*}
where $D/Dt$ is the material derivative, defined as:
\begin{linenomath*}
\begin{equation}
\frac{D}{Dt}\equiv \frac{\partial}{\partial t} + \mathbf{U}\cdot\nabla \;.
\end{equation}
\end{linenomath*}
Here, $\mathbf{U}$ represents the three-dimensional wind vector with its horizontal and vertical components $\mathbf{v}$ and $w$, $\nabla$ is the three-dimensional gradient operator, $\theta \equiv T(p_0/p)^{{R_d/c_p}}$ denotes the potential temperature with $p_0$ as a constant reference pressure and $c_p$ as the specific heat of dry air at constant pressure, $\Pi\equiv (p/p_0)^{{R_d/c_p}}$ is the Exner function, $\nabla\cdot$ is the three-dimensional divergence operator, $\mathbf{R}$ is the radiative heat flux vector, $\mathbf{F}_t$ is the enthalpy flux vector due to conduction and unresolved motions\footnote{Note that this term encompasses any contribution from a convective parameterization, if one is employed.  Additionally, ModelE3 omits the enthalpy flux carried by falling precipitation.}, $L_v$ is the latent heat of vaporization, $\mathcal{C}$ is the net condensation rate (condensation minus evaporation) per unit mass, $L_f$ is the latent heat of fusion, $\mathcal{F}$ is the net freezing rate (freezing minus melting) per unit mass, $L_s\equiv L_v+L_f$ is the latent heat of sublimation, $\mathcal{D}$ is the net deposition rate (deposition minus sublimation) per unit mass, and $\delta$ represents the rate of viscous dissipation of kinetic energy per unit mass, known as frictional heating. A more detailed description of frictional heating is provided in \ref{subsec:total}. By applying the product rule to Eq.~\ref{eq:potential_temp} and utilizing the ideal gas law from Eq.~\ref{eq:ideal_gas}, we obtain
\begin{linenomath*}
\begin{equation}\label{eq:enthalpy}
\frac{D}{Dt}\kirb{c_pT}  -\frac{1}{\rho}\frac{Dp}{Dt} = -\frac{1}{\rho}\nabla\cdot\kirb{\mathbf{R}+\mathbf{F}_t} +L_v\mathcal{C}+L_f\mathcal{F} + L_s\mathcal{D} + \delta\,.
\end{equation}
\end{linenomath*}

The conservation of water vapor is governed by the following equation:
\begin{linenomath*}
\begin{equation}\label{eq:qv}
\frac{Dq_v}{Dt} = -\mathcal{C}-\mathcal{D}-\frac{1}{\rho}\nabla\cdot \mathbf{F}_v\;, 
\end{equation}
\end{linenomath*}
where $q_v$ is the water vapor mixing ratio, and $\mathbf{F}_v$ represents the water vapor flux vector due to unresolved motions. Similarly, the conservation of ice is expressed as:
\begin{linenomath*}
\begin{equation}\label{eq:qi}
\frac{Dq_i}{Dt} = \mathcal{F}+\mathcal{D}-\frac{1}{\rho}\nabla\cdot\kirb{\mathbf{F}_i-\mathbf{P}_i} \;,
\end{equation}
\end{linenomath*}
where $q_i$ represents the ice mixing ratio, encompassing ice in clouds and snowfall, $\mathbf{F}_i$ is the ice flux vector due to unresolved motions excluding snowfall, and $\mathbf{P}_i$ is the snowfall flux vector. It should be noted that $\mathbf{P}_i$ is defined to be positive for downward fluxes, contrasting with the convention for the other flux vectors which are positive for upward fluxes. 

In the equations presented thus far, we have excluded chemical processes such as the oxidation of atmospheric methane in the stratosphere and at higher altitudes because their impact on the column MSE budget is so minor that it is not discernible in any of the analyses conducted in this study.

By adding $D\phi/Dt \equiv gw$ to Eq.~\ref{eq:enthalpy}, the dry static energy (DSE) equation is derived, as detailed in \ref{sec:dse}.  By additionally adding $L_v$ times $\text{Eq.~\ref{eq:qv}}$ and subtracting $L_f$ times $\text{Eq.~\ref{eq:qi}}$, we obtain:
\begin{linenomath*}
\begin{equation}\label{eq:fmse}
\frac{Dh}{Dt} = - \frac{1}{\rho}\nabla\cdot \kirb{\mathbf{R} + L_f\mathbf{P}_i  +\mathbf{F}_h} +\epsilon +\delta \;,
\end{equation}
\end{linenomath*}
where $h \equiv c_pT + \phi + L_vq_v - L_fq_i$ is the MSE (often referred to as frozen MSE), $\phi\equiv gz$ is the geopotential energy with the gravitational acceleration $g$ and the height $z$, $\mathbf{F}_h\equiv\mathbf{F}_t + L_v\mathbf{F}_v-L_f\mathbf{F}_i$ is the MSE flux vector due to conduction and unresolved motions excluding snowfall. 

We define
\begin{linenomath*}
\begin{eqnarray}
\epsilon &\equiv& \frac{1}{\rho}\frac{Dp}{Dt} + gw \label{eq:epsilon_1} \\
&=& \frac{1}{\rho}\frac{\partial p}{\partial t} + \frac{1}{\rho}\mathbf{v}\cdot\nabla_z p + \frac{w}{\rho}\kirb{\frac{\partial p}{\partial z} + \rho g} \label{eq:epsilon_2} \\  
& \equiv & \frac{1}{\rho}\frac{\partial p}{\partial t} + \widetilde{\epsilon} \;, \label{eq:epsilon_3}
\end{eqnarray}
\end{linenomath*} 
where $\nabla_z$ is the horizontal gradient operator at constant height. The last two terms of Eq.~\ref{eq:epsilon_2} are combined and denoted as:
\begin{linenomath*}
\begin{equation}\label{eq:epsilon_tilde}
\widetilde{\epsilon} \equiv \frac{1}{\rho}\mathbf{v}\cdot\nabla_z p + \frac{w}{\rho}\kirb{\frac{\partial p}{\partial z} + \rho g} \;,
\end{equation}
\end{linenomath*}
which represents the work performed against pressure-gradient and buoyancy forces. Under hydrostatic balance, the last term vanishes, leaving
\begin{linenomath*}
\begin{equation}\label{eq:epsilon_tilde_hydro}
\widetilde{\epsilon} \equiv \frac{1}{\rho}\mathbf{v}\cdot\nabla_z p  \qquad \text{(hydrostatic balance)} \;.
\end{equation}
\end{linenomath*}

When $\widetilde{\epsilon}$ is averaged over a grid box, sub-grid covariance terms appear.  In models that employ a convective parameterization, whether the dynamics are hydrostatic or non-hydrostatic, the buoyancy contribution in Eq.~\ref{eq:epsilon_tilde} is the dominant source of these covariances.  With straightforward algebra, however, the covariance terms can be absorbed into the turbulent-flux divergence $\nabla \cdot \mathbf{F}_t$ and the frictional-heating term $\delta$; therefore, no separate treatment is needed when deriving the budget equations.  Consequently, all variables in Eqs.~\ref{eq:epsilon_tilde}–\ref{eq:epsilon_tilde_hydro} ($\rho$, $p$, $\mathbf{v}$, $w$) can be regarded as grid-resolved (grid-averaged) quantities.  The detailed rationale for this grouping is provided in \ref{subsec:hydrostatic}.

In an accelerating parcel, $\widetilde{\epsilon}$ is negative (detailed in \ref{subsec:total}), and thus acts as a net sink of enthalpy or MSE. This sink can damp convective activity during the development of convective systems. Accurately representing this effect in numerical models is therefore essential.

By multiplying Eq.~\ref{eq:fmse} with $\rho$ and applying the mass conservation equation,
\begin{linenomath*}
\begin{equation}\label{eq:mass_conservation}
\frac{\partial\rho}{\partial t} + \nabla\cdot\kirb{\rho \mathbf{U}} = 0\;,
\end{equation}
\end{linenomath*}
we derive the flux-divergence form of the MSE budget:
\begin{linenomath*}
\begin{equation}\label{eq:flux_div_mse}
 \frac{\partial }{\partial t}\kirb{\rho h} = - \nabla\cdot\kirb{\rho h\mathbf{U}} -\nabla\cdot \kirb{\mathbf{R}+L_f\mathbf{P}_i + \mathbf{F}_h} + \rho\epsilon + \rho\delta\;.
\end{equation}
\end{linenomath*}
Upon integrating with respect to $z$, we obtain
\begin{linenomath*}
\begin{equation}\label{eq:column_mse}
\frac{\partial \kiab{h}}{\partial t} = -\nabla_z\cdot\kiab{h\mathbf{v}} +R + L_fP_{i,s} + L_vE + H + \kiab{\epsilon} + \kiab{\delta}\;.
\end{equation}
\end{linenomath*}
A detailed derivation from Eq.~\ref{eq:flux_div_mse} to Eq.~\ref{eq:column_mse} is presented in \ref{sec:from_flux_column}. Here, the mass-weighted vertical (or column) integration from the surface height $z_s$ to the top-of-the-atmosphere height $z_t$ is defined as
\begin{linenomath*}
\begin{equation}
\kiab{X} \equiv \int_{z_s}^{z_t} X\rho \ud z \;, 
\end{equation}
\end{linenomath*}
$\nabla_z\cdot$ denotes the horizontal divergence operator for a two-dimensional vector, $R \equiv R_s-R_t$ represents the column radiative heating, with $R_s$ and $R_t$ being the upward net radiative heat fluxes at the surface and the top of the atmosphere, respectively, and $P_{i,s}$ denotes the surface snowfall, defined as positive for a downward flux. The surface value of the vertical component of $\mathbf{F}_h$ is given by $L_vE+H$, where $E$ is the surface evaporation and $H$ is the surface sensible heat flux.

In this derivation, the horizontal components of $\mathbf{R}$, $\mathbf{P}_i$, and $\mathbf{F}_h$ are assumed to be zero, as is the case in ModelE3. However, this assumption is not universally applicable across all models. High-resolution cloud resolving models often retain a non-zero horizontal component of $\mathbf{F}_{h}$, denoted $\mathbf{F}_{h,h}$ (see \citeA{guichard_short_2017} and references therein), and must therefore account for its contribution to the MSE budget. Even some hydrostatic climate models---such as the NCAR Community Atmosphere Model \cite<CAM~4.0;>[]{neale_description_nodate}---include explicit horizontal diffusion of temperature. In such cases, it is necessary to incorporate $-\kiab{\rho^{-1}\nabla_z\cdot \mathbf{F}_{h,h}}$ into the RHS of Eq.~\ref{eq:column_mse}. Analogous corrections are also required for $\mathbf{R}$ and $\mathbf{P}_i$ when their full three-dimensional contributions are considered.

Often, we simplify the column MSE budget equation by neglecting terms such as $L_fP_{i,s}$, $\kiab{\epsilon}$, and $\kiab{\delta}$ in Eq.~\ref{eq:column_mse} \cite<e.g.,>[]{yanai_determination_1973, neelin_modeling_1987, bretherton_energy-balance_2005, maloney_moist_2009, wing_convective_2017}, leading to the following approximation:
\begin{linenomath*}
\begin{equation}\label{eq:column_fmse_approx}
\frac{\partial \kiab{h}}{\partial t} \simeq -\nabla_z\cdot\kiab{h\mathbf{v}} +R +  L_vE + H \;.
\end{equation}
\end{linenomath*}

Given that the surface height remains constant over time in ModelE3 (i.e., $\partial z_s/\partial t = 0$) and $\rho=0$ at the top of the atmosphere, we derive, from the ideal gas law (Eq.~\ref{eq:ideal_gas}):
\begin{linenomath*}
\begin{equation}\label{eq:column_epsilon}
\kiab{\epsilon} = \int_{z_s}^{z_t} \frac{\partial (\rho R_dT)}{\partial t} \ud z + \int_{z_s}^{z_t}\widetilde{\epsilon}\rho \ud z = \frac{\partial}{\partial t}\kiab{R_dT} + \kiab{\widetilde{\epsilon}}\;.
\end{equation}
\end{linenomath*}
Consequently, Eq.~\ref{eq:column_mse} can be reformulated as:
\begin{linenomath*}
\begin{equation}\label{eq:column_mse2}
\frac{\partial}{\partial t}\kiab{h-R_dT} = -\nabla_z\cdot\kiab{h\mathbf{v}} +R + L_fP_{i,s} + L_vE + H + \kiab{\widetilde{\epsilon}} + \kiab{\delta}\;,
\end{equation}
\end{linenomath*}
where $h-R_dT = c_vT + \phi + L_vq_v - L_fq_i$, given that $c_p \equiv c_v + R_d$.

Finally, we slightly modify Eq.~\ref{eq:column_mse2} to remove $\phi$ from the Eulerian tendency term. This adjustment is necessary because $\phi$ is not a native quantity in hydrostatic models; it is diagnostically computed from temperature and pressure. Removing $\phi$ facilitates the accurate computation of the column MSE budget, as discussed in Section~\ref{subsec:before_after}. By applying the hydrostatic balance and the ideal gas law, and integrating by parts under the assumption that $p=0$ at the top of the atmosphere, we derive:
\begin{linenomath*}
\begin{equation}
\kiab{\phi} = z_sp_s + \kiab{R_dT}\;.
\end{equation}
\end{linenomath*} 
Therefore, Eq.~\ref{eq:column_mse2} in hydrostatic models can be expressed as:
\begin{linenomath*}
\begin{equation}\label{eq:column_mse3}
\frac{\partial}{\partial t}  \kirb{ \langle\widetilde{h} \rangle +z_sp_s} = -\nabla_z\cdot\kiab{h\mathbf{v}} +R + L_fP_{i,s} + L_vE + H + \kiab{\widetilde{\epsilon}} + \kiab{\delta}\;,
\end{equation}
\end{linenomath*}
where $\widetilde{h}\equiv c_pT + L_vq_v - L_fq_i$ is referred to as the (frozen) moist enthalpy, and $p_s$ denotes the surface pressure.

Although some studies analyze the total energy budget---including both MSE and kinetic energy---this study employs the MSE budget alone because it affords greater precision and conceptual clarity for our analysis. The underlying rationale, along with the full derivation of the total‐energy equation, is given in \ref{subsec:total}.

\subsection{Expanding into Advective Forms}
In this subsection, we express the MSE budget equation in its advective forms within two vertical coordinate systems often utilized in MSE studies. By expanding the material derivative within the $z$-coordinate system, we can reformulate Eq.~\ref{eq:fmse} as follows:
\begin{linenomath*}
\begin{equation}\label{eq:fmse_z}
\kirb{\frac{\partial h}{\partial t}}_z = -\mathbf{v}\cdot\nabla_z h - w\frac{\partial h}{\partial z} -\frac{1}{\rho}\frac{\partial}{\partial z}\kirb{R_v + L_fP_{i,v}+ F_{h,v} } + \epsilon + \delta \;,
\end{equation}
\end{linenomath*}
where the subscript $z$ is used to emphasize that the Eulerian tendency is computed in the $z$-coordinate system. Here, $R_v$, $P_{i,v}$, and $F_{h,v}$ denote the vertical components of $\mathbf{R}$, $\mathbf{P}_i$, and $\mathbf{F}_h$, respectively. Consistent with our earlier discussion, their horizontal components are assumed to be zero.  Performing a column integration of Eq.~\ref{eq:fmse_z} yields:
\begin{linenomath*}
\begin{equation}\label{eq:column_mse_z}
\kiab{\kirb{\frac{\partial h}{\partial t}}_z} = -\kiab{\mathbf{v}\cdot\nabla_z h} - \kiab{w\frac{\partial h}{\partial z}} +R + L_fP_{i,s} + L_vE + H + \kiab{\epsilon} + \kiab{\delta}\;.
\end{equation}
\end{linenomath*}

We now transition from the $z$-coordinate system to the hydrostatic pressure coordinate system (referred to as the $p$-coordinate system). 
Referencing Eqs.~3.3 and 3.4 from \citeA{kasahara_various_1974}, we establish the following relationships:
\begin{linenomath*}
\begin{equation}\label{eq:change_coord}
-\frac{1}{\rho}\frac{\partial X}{\partial z} = g\frac{\partial X}{\partial p}\;, \quad \text{and } \quad \kirb{\frac{\partial X}{\partial c}}_z = \kirb{\frac{\partial X}{\partial c}}_p + \rho\kirb{\frac{\partial X}{\partial p}}\kirb{\frac{\partial \phi}{\partial c}}_p\;, 
\end{equation}
\end{linenomath*}
where $c = x,y,t$.
By applying Eq.~\ref{eq:change_coord} to Eq.~\ref{eq:fmse_z}, along with the relation:
\begin{linenomath*}
\begin{equation}
w \equiv \frac{Dz}{Dt} = \kirb{\frac{\partial z}{\partial t}}_p + \mathbf{v}\cdot\nabla_p z + \omega_p\frac{\partial z}{\partial p} \;,
\end{equation}
\end{linenomath*}
we obtain
\begin{linenomath*}
\begin{equation}\label{eq:fmse_p}
\kirb{\frac{\partial h}{\partial t}}_p = - \mathbf{v}\cdot\nabla_p h - \omega_p\frac{\partial h}{\partial p} + g\frac{\partial}{\partial p}\kirb{R_v + L_fP_{i,v}+F_{h,v}} + \epsilon + \delta \;,
\end{equation}
\end{linenomath*}
where
\begin{linenomath*}
\begin{equation}\label{eq:epsilon_hydro2}
\epsilon = \kirb{\frac{\partial \phi}{\partial t}}_p + \mathbf{v}\cdot\nabla_p \phi\  \;,
\end{equation}
\end{linenomath*}
which is equivalent to Eq.~\ref{eq:epsilon_1} in the $p$-coordinate system. Here, the subscript $p$ signifies that the Eulerian tendency is evaluated in the $p$-coordinate system, $\nabla_p$ denotes the horizontal gradient operator at constant pressure, and $\omega_p$ is the vertical pressure velocity.

The column integration in the $p$-coordinate system can be formulated through a change of variables as:
\begin{linenomath*}
\begin{equation}
\kiab{X} \equiv \int_{z_s}^{z_t}X(z)\rho(z) \ud z = \frac{1}{g}\int_0^{p_s}X(p)\ud p \;.
\end{equation}
\end{linenomath*}
Applying this column integration to Eq.~\ref{eq:fmse_p}, we derive the column MSE budget equation in its advective form within the $p$-coordinate system:
\begin{linenomath*}
\begin{equation}\label{eq:column_mse_p}
\kiab{ \kirb{\frac{\partial h}{\partial t}}_p  } = -\kiab{\mathbf{v}\cdot\nabla_p h } -\kiab{ \omega_p\frac{\partial h}{\partial p}} + R + L_fP_{i,s} + L_vE + H  + \kiab{\epsilon} + \kiab{\delta} \;.
\end{equation}
\end{linenomath*}
Utilizing the Leibniz integral rule, we further deduce:
\begin{linenomath*}
\begin{equation}\label{eq:tend_mse_p}
\kiab{ \kirb{\frac{\partial h}{\partial t}}_p  }  = \frac{\partial \kiab{h}}{\partial t} - \frac{1}{g}h_s\frac{\partial p_s}{\partial t} \;.
\end{equation}
\end{linenomath*}
A comparison between Eq.~\ref{eq:column_mse} and Eq.~\ref{eq:column_mse_p}, along with Eq.~\ref{eq:tend_mse_p}, yields:
\begin{linenomath*}
\begin{equation}\label{eq:flux_advective_mse}
\nabla_z\cdot\kiab{h\mathbf{v}} = -\frac{1}{g}h_s\frac{\partial p_s}{\partial t} + \kiab{\mathbf{v}\cdot\nabla_p h } + \kiab{ \omega_p\frac{\partial h}{\partial p}} \;.
\end{equation}
\end{linenomath*}
This relation can also be directly derived by applying the Leibniz integral rule to $\smash{\nabla_z\cdot\kiab{h\mathbf{v}}}$, as detailed in \ref{sec:derivation_advective}.

Equations~\ref{eq:fmse_p} and \ref{eq:column_mse_p} can be rearranged by merging the term $\epsilon$ into the Eulerian tendency and the horizontal advection, as follows:
\begin{linenomath*}
\begin{equation}\label{eq:enthalpy_advection_form}
\kirb{\frac{\partial \widetilde{h}}{\partial t}}_p = - \mathbf{v}\cdot\nabla_p \widetilde{h} - \omega_p\frac{\partial h}{\partial p} + g\frac{\partial}{\partial p}\kirb{R_v + L_fP_{i,v}+F_{h,v}}  + \delta \;,
\end{equation}
\end{linenomath*}
\begin{linenomath*}
\begin{equation}\label{eq:column_mse_p2}
\kiab{ \kirb{\frac{\partial \widetilde{h}}{\partial t}}_p  } = -\kiab{\mathbf{v}\cdot\nabla_p \widetilde{h} } -\kiab{ \omega_p\frac{\partial h}{\partial p}} + R + L_fP_{i,s} + L_vE + H  + \kiab{\delta} \;.
\end{equation}
\end{linenomath*}
In the $p$-coordinate system, the accurate formulation of the MSE budget equation can be achieved by substituting $h$ with $\widetilde{h}$ in both the Eulerian tendency and the horizontal advection terms.

\section{Description of ModelE3 and Data Analyzed}\label{sec:model}
This section provides a brief overview of the ModelE3 configurations relevant to this study. For further description of the model's physics, readers are referred to \citeA{cesana_snow_2021} and \citeA{stanford_observed_2023}. Note that an alternative version of ModelE3, which utilizes a different dynamical core, is also available. However, the methodology introduced in this study is independent of the choice of dynamical core. The version of ModelE3 used in our analysis employs a 30-minute dynamics integration time step, a horizontal resolution of 2$^\circ\times$2.5$^\circ$ in latitude and longitude, and 110 vertical levels in a hybrid coordinate system. This system uses a sigma coordinate up to 150 hPa, transitioning to a pressure coordinate at higher altitudes. Horizontal velocities are placed on the Arakawa-B grid. Tracers such as potential temperature and humidity are advected using the quadratic upstream scheme \cite<QUS;>[]{prather_numerical_1986}. This scheme represents the sub-grid distribution of tracers as second-order polynomials in three dimensions, thereby enhancing the effective resolution of the tracer field. The simulation was initialized on December 1, 1999, with an initial one-month period for model spin-up, followed by a one-year run. Data spanning from January 1 to December 31, 2000, were analyzed for the results presented in this study.

To illustrate the magnitude of column MSE budget residuals computed from model outputs, we analyzed both raw and postprocessed data. The raw data are provided at the model's highest resolution: a temporal frequency of every 30 minutes and spanning 110 vertical levels in the model's native coordinate system (referred to as the m-coordinate system). The postprocessing includes vertical interpolation and daily averaging, with logarithmic interpolation transforming data from the m-coordinate to the pressure-coordinate system at 32 specified pressure levels: 1000, 975, 950, 925, 900, 875, 850, 825, 800, 775, 750, 700, 650, 600, 550, 500, 450, 400, 350, 300, 250, 225, 200, 175, 150, 125, 100, 70, 50, 30, 20, and 10 hPa.

We have also integrated new code within the model to accurately compute the column MSE budget, focusing particularly on its column flux divergence, as elaborated in Section~\ref{subsec:before_after}.


\section{Residual Computed from Postprocessed Data}\label{sec:residual}
\begin{figure}[!ht] 
\includegraphics[width=\textwidth]{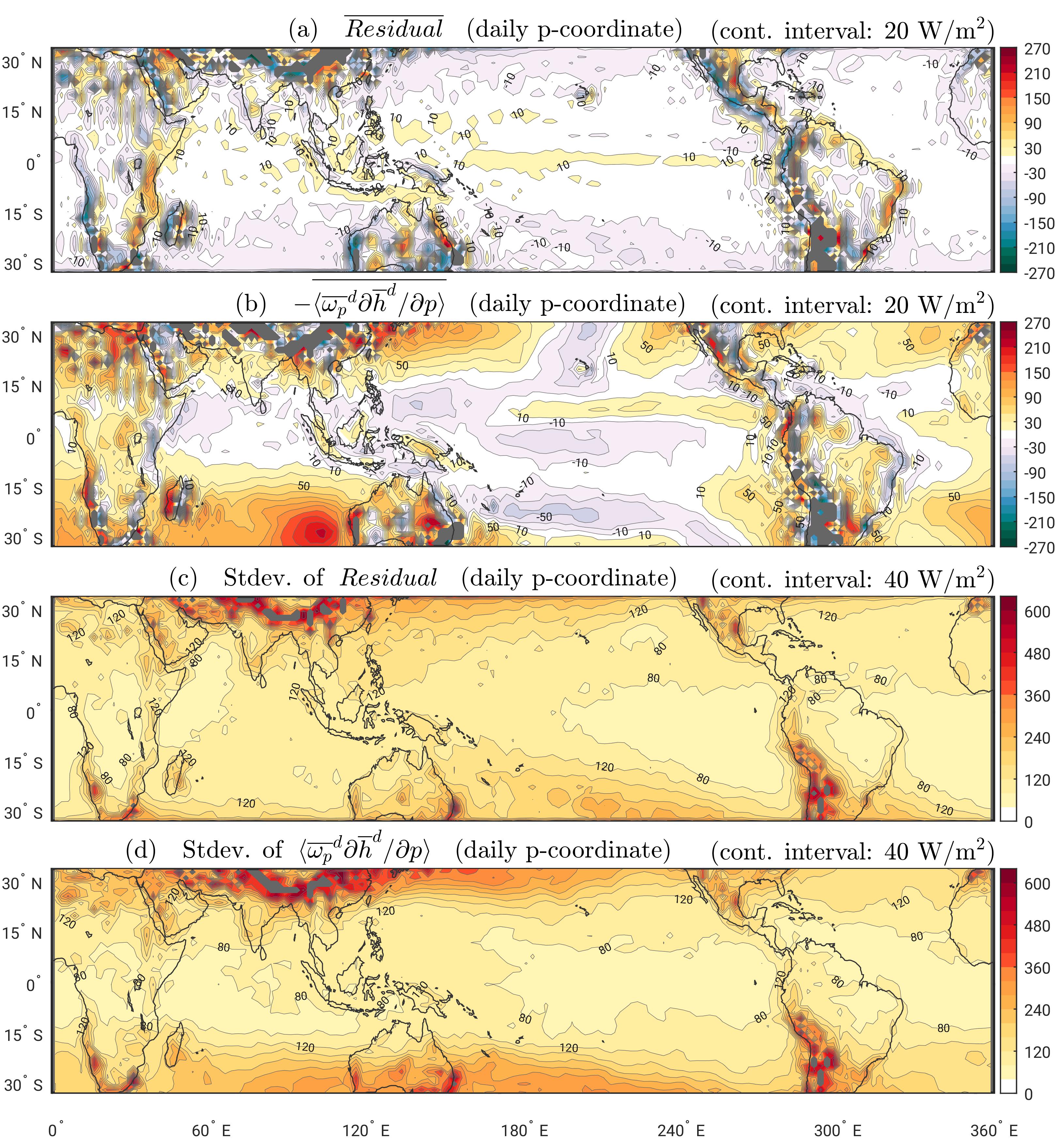} 
\caption{Annual mean (a, b) and standard deviation (c, d) of the column MSE budget residual and the vertical advection, $-\kiab{\omega_p\partial h/\partial p}$, calculated using daily averaged variables in the $p$-coordinate system. Contour intervals are specified in each panel, with regions of saturated contours masked in gray.}\label{fig:res_pp} 
\end{figure}

We start with the demonstration of the magnitude of the column MSE budget residual, computed using the postprocessed data which involve the vertical interpolation and the daily average of each variable. 
In Fig.~\ref{fig:res_pp}, we display the annual mean and the daily standard deviation of this budget residual. It is computed by subtracting the RHS from the LHS of Eq.~\ref{eq:column_mse_p} (or Eq.~\ref{eq:column_mse_p2}), where the Eulerian tendency and the advective terms are calculated using the postprocessed data and second-order centered differences in time and space. To facilitate a comparison of its magnitude, we additionally display, in Figs.~\ref{fig:res_pp}(b) and (d), the column-integrated vertical MSE advection, $-\langle{\omega_p\partial h/\partial p}\rangle$, computed using the postprocessed $\omega_p$ and $h$. The vertical advection was chosen as a reference because of its relevance to the ability of tropical circulations to export energy out of the tropics and of convection to strengthen and deepen \cite<e.g.,>[]{neelin_modeling_1987, raymond_mechanics_2009, inoue_gross_2015}.

Figures~\ref{fig:res_pp}(a) and (b) reveal that the annual mean of the residual often ranges between 10--30~W~m$^{-2}$ over tropical oceans, a magnitude comparable to that of the vertical advection in these regions. Notably, the vertical advection shows minimal values within the range of $\pm$10 W~m$^{-2}$ over the Indian Ocean, to the extent that its sign could be reversed by the residual. Over subtropical oceans, the residual is predominantly negative, exhibiting a slightly greater magnitude than that observed in the deep tropical oceans. Across continental areas, the magnitude of the residual sharply increases, particularly in the vicinity of topographical features like the Andes Mountains and the Himalayas, where it exceeds the contour limits. The significant size of the annual mean residual, comparable to that of the vertical advection, complicates the task of discerning signals from the vertical advection \cite<e.g.,>[]{back_geographic_2006}.

Regarding daily variability, the significance of the residual becomes even more pronounced. Figures~\ref{fig:res_pp}(c) and (d) illustrate that the standard deviation of the residual  generally surpasses that of the vertical advection in the deep tropical oceans. This substantial magnitude of the residual's fluctuation complicates the evaluation of the vertical advection's influence on convective variability \cite<e.g.,>[]{kim_propagating_2014, ren_intercomparison_2021}. 

It should be noted that the magnitude of the residual obtained from the approximated column MSE budget equation (Eq.~\ref{eq:column_fmse_approx})---where terms $\kiab{\epsilon}$, $L_fP_{i,s}$, and $\kiab{\delta}$ are omitted---is nearly identical to that calculated from the complete equation (refer to Fig.~S1 in Supporting Information). Specifically, the domain-averaged residual derived from the complete equation (in Fig.~\ref{fig:res_pp}(a)) is $-5.615$ W~m$^{-2}$, while that derived from the approximated equation is $-5.832$ W~m$^{-2}$ (in Fig.~S1(a)). The difference between them is negligible. This observation suggests that the simplification in the column MSE budget equation does not significantly influence the budget closure. The roots of these residuals are, instead, intricately linked to the interplay of the factors discussed in the following section.

\section{Challenges in Calculating the MSE Budget and Causes of Large Residuals}\label{sec:cause}
In this section, we identify seven interconnected reasons why the MSE budget fails to close.  Although the examples are drawn from ModelE3, most of the limitations apply equally to other climate models; additional factors may of course exist beyond those identified here.  The common thread linking all seven issues is the difficulty of diagnosing the transport term \textit{a posteriori} from model output that was never designed for strict MSE conservation.

\noindent\textbf{Reason 1: Discrepancy between Continuous and Discrete Calculus.}
One fundamental reason for the residual in the MSE budget computation is the fact that continuous calculus properties, such as the product and chain rules, do not hold when applied to discretized configurations.  For example, the product rule
\begin{linenomath*}
\begin{equation}\label{eq:product}
\frac{d(AB)}{dl} \;=\; A\,\frac{dB}{dl} \;+\; B\,\frac{dA}{dl} \; , \qquad  l\in\{x,y,z,t\}, \;
\end{equation}
\end{linenomath*}
holds analytically but not precisely on a numerical grid.  This mismatch motivates a variety of advection schemes, each designed to minimize the numerical errors in its own way \cite<elegantly summarized by>{souza_flux-differencing_2023}.  

In the MSE budget, the failure of the product rule invalidates the continuous identity that links the potential-temperature and enthalpy forms:
\begin{linenomath*}
\begin{equation}\label{eq:theta_enth_identity}
c_p\Pi \kirb{\frac{\partial (\rho \theta)}{\partial t} + \nabla\cdot (\rho\theta\mathbf{U}) } = \frac{\partial (\rho c_pT)}{\partial t} + \nabla\cdot \kirb{\rho c_pT\mathbf{U}}-\kirb{ \frac{\partial p}{\partial t} + \mathbf{U}\cdot\nabla p } \;.
\end{equation}
\end{linenomath*}
Once the governing equations are discretized, Eq.~\ref{eq:theta_enth_identity} no longer holds exactly. ModelE3 integrates $\theta$ prognostically, so its advection operator conserves $\theta$ by construction.  Applying that same operator to enthalpy or MSE, however, inherits the broken identity and leaves a non-zero residual.  Achieving a closed MSE budget therefore demands an advection scheme tailored to MSE itself; the implementation is described in Section~\ref{subsec:before_after}.

The product rule issue may be particularly pronounced in scenarios with steep pressure and temperature gradients. This can result in larger residuals in the MSE budget in areas with intense gradients, such as regions of high topography, in the presence of baroclinic storms, and in the vicinity of tropical cyclones. Additionally, coarse resolution can further exacerbate the failure of continuous calculus properties.

\noindent \textbf{Reason 2: Impact of Mass Filtering on MSE Transport.}
The version of ModelE3 used in this study applies numerical filters to the mass field to mitigate computational modes. When this mass filter is applied, it alters the column mass and consequently changes the pressure, which in hydrostatic models is vertically integrated mass. This adjustment affects the Exner function $\Pi$, thereby altering the temperature, which is expressed as $T=\Pi\theta$ where $\theta$ is conserved. This effect---adiabatic heating or cooling resulting from mass changes due to filtering---can be physically interpreted as vertical advection of DSE and consequently MSE. Mathematically, this effect should be included in the column DSE flux divergence. However, this effect would not be perfectly reconciled with the column DSE flux divergence in discretized models, leading to discrepancies. Therefore, reconstructing this form of DSE (and MSE) transport using standard model outputs is likely unattainable, resulting in a discrepancy between the actual MSE transport and the diagnostically computed values.

\noindent \textbf{Reason 3: Challenges in Reconstructing Flux Divergence Using Standard Outputs.}
Tracer transport is often diagnostically computed using a flux-divergence form (like $\nabla_z\cdot\kiab{\; \mathbf{v}}$) rather than an advective form, primarily to ensure consistency with a model's physics that transports tracers in a flux-divergence form. However, this approach presents a significant issue: accurately reconstructing flux divergence from model outputs is generally unattainable. This difficulty arises because the divergence operator in numerical models often cannot be represented by a simple equation or expression. Despite this, attempts are frequently made to reconstruct flux divergence using a centered difference method, which leads to significant errors. This issue has also been observed in reanalysis data, which archive both column flux and column flux divergence of moisture but these cannot be reconciled with a centered difference method \cite{seager_diagnostic_2013}.

Because the flux–divergence formulation amplifies even minute errors in the estimated mass-flux divergence, any inconsistency quickly propagates through the budget (detailed in Section~\ref{subsec:direct_method}). To mitigate this problem, several studies \cite<e.g.,>[]{trenberth_climate_1991, trenberth_using_1997, peters_analysis_2008} impose \textit{a posteriori} mass corrections on the divergent wind. Such corrections, however, assume that the error field follows a prescribed---usually barotropic---vertical structure and therefore may misrepresent the true three-dimensional error distribution. Consequently, diagnostically derived mass-flux divergences often leave sizeable residuals in the MSE budget, as examined in Section~\ref{subsec:direct_method}.

\noindent\textbf{Reason 4: Inconsistency of advective form with the model's flux‐form transport.}  
Given the obstacle above, one might prefer the advective form (Eqs.~\ref{eq:column_mse_p} and \ref{eq:column_mse_p2}) over the flux‐divergence form (Eqs.~\ref{eq:column_mse}, \ref{eq:column_mse2}, and \ref{eq:column_mse3}) when evaluating the transport term.  However, ModelE3 (and most models) actually transport tracers in flux form.  Using the advective representation therefore mismatches the model's native discretization, again producing a systematic residual in the MSE budget. Section~\ref{subsec:direct_method} contrasts the two approaches (advective vs. flux-divergence forms) and shows which incurs the smaller error in practice.

\begin{figure}[!ht] 
\includegraphics[width=\textwidth]{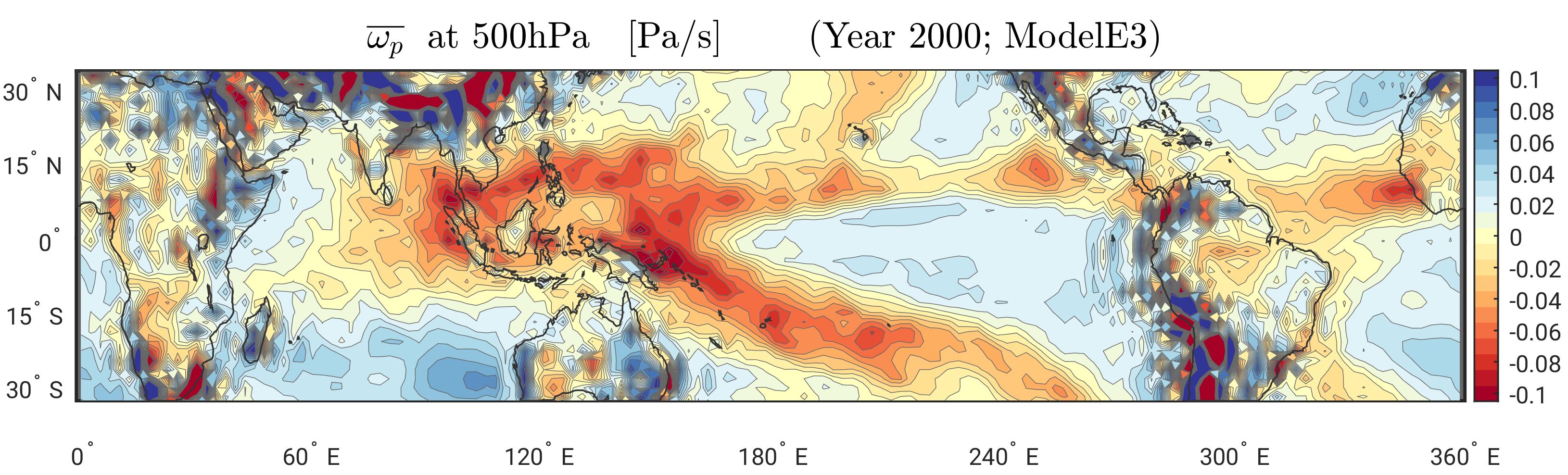} 
\caption{Annual mean of $\omega_p$ at 500~hPa from ModelE3 for the year 2000.}\label{fig:omega} 
\end{figure}

\noindent \textbf{Reason 5: Numerical Errors in the Output of $\boldsymbol{\omega_p}$.}
The vertical wind, $\omega_p$ or $w$, is not a native quantity in hydrostatic models and is therefore computed diagnostically. This diagnostically computed vertical wind often introduces errors into MSE budget computations, particularly when used with the advective form equations. The computation of vertical wind is prone to contamination by numerical errors, especially around topographic features. In fact, the $\omega_p$ output from ModelE3's diagnostic code is heavily contaminated by numerical errors over continental regions. Figure~\ref{fig:omega} shows an example at
500 hPa where $\omega_p$ may attain its peak value. This contamination becomes the primary source of errors in MSE budget computations in those areas, as explored in Section~\ref{sec:results}. This issue may arise from deficiencies in ModelE3's diagnostic code, and there is potential for improvement by modifying the code to more effectively mitigate the effects of topography.

\noindent  \textbf{Reason 6: Timing of Model Output Harvesting.}
Another potential reason for the MSE budget residual in ModelE3, and models more generally, may stem from the timing of model output harvesting. In ModelE3, each prognostic variable undergoes a series of integration schemes: turbulence $\rightarrow$ cloud $\rightarrow$ radiation $\rightarrow$ dynamics schemes (see also Fig.~\ref{fig:flow} in Section~\ref{sec:method}). Throughout this sequence, the variable is sequentially updated, each stage adding its corresponding tendency. 
Notably, the harvesting of each variable for output occurs immediately following the cloud scheme, resulting in a harvested output that differs from the input used in the dynamics scheme, which follows the radiation scheme.
Consequently, the quantities harvested as model outputs are not identical to those being advected within the model. This discrepancy suggests that any attempt to compute the advective tendencies directly from standard model outputs, regardless of their resolution or coordinate system, 
is susceptible to inaccuracies, potentially leading to errors in the calculated advective tendencies.

\noindent \textbf{Reason 7: Vertical Interpolation and Temporal Averaging.}
Model outputs are typically postprocessed---most commonly by vertically interpolating fields to the $p$-coordinate and applying temporal averaging---to facilitate analysis. These steps, however, can introduce additional errors into the MSE budget. First, the $p$-coordinate is not the model's native vertical coordinate, so mapping prognostic variables onto fixed-$p$ levels inevitably introduces spurious errors. Second, the vertical interpolation is performed offline rather than inline with tracer advection: the advection scheme evolves tracers on the native vertical coordinate, experiencing the ``true'' time-evolving pressure of each layer, whereas the interpolated fields are referenced to a different vertical grid than that experienced by the advection scheme. This mismatch can bias budget terms and disrupt the exact equivalence between flux-form and advective-form diagnostics. Third, temporal averaging compounds these issues, thereby aliasing fluctuations in pressure surfaces and advective fluxes into the averaged fields. Collectively, these postprocessing steps can degrade MSE budget closure and exacerbate other error sources.

These factors interact rather than act in isolation. For example, the product-rule violation (Reason~1) is aggravated by asynchronous output (Reason~6); using the advective form (Reason~4) amplifies the mismatch in Eq.~\ref{eq:theta_enth_identity} (Reason~1); and so forth.  

The MSE transport term is inherently prone to small errors, mainly because it results from the cancellation of two large terms \cite<e.g.,>[]{Neelin_2008_10, raymond_mechanics_2009}. This sensitivity presents a significant challenge to the accuracy of the computation. Essentially, the key goal is to maintain errors that are substantially smaller than the actual signals. However, as detailed in Section~\ref{subsec:direct_method}, the errors stemming from the unavoidable reasons mentioned above are significant enough to obscure the signal itself, complicating the accurate computation of the MSE budget.

Note that the factors influencing the MSE budget residual are not consistent across all models.  For instance, some dynamical core options in CAM~4.0 \cite{neale_description_nodate} use enthalpy (similar to Eq.~\ref{eq:enthalpy}) as the prognostic temperature variable, effectively bypassing the product rule issue. In the same vein, the Nonhydrostatic Icosahedral Atmospheric Model \cite<NICAM;>[]{satoh_nonhydrostatic_2008} utilizes internal energy as its prognostic variable, also avoiding the product rule issue. Models such as the System for Atmospheric Modeling \cite<SAM;>[]{khairoutdinov_cloud_2003, khairoutdinov_global_2022} implement static energies as their prognostic variables, thereby ensuring precise MSE conservation. Additionally, different models employ various advection schemes, numerical filtering, grid resolutions, and strategies for output harvesting, all contributing to the distinct nature of each model's MSE budget residual. Consequently, the magnitude of the MSE budget residual may be highly dependent on the specific physics and configurations unique to each model. Observing smaller residuals in some models doesn't inherently imply superiority; rather, it could simply be a reflection of their distinct structural designs. 

Fundamentally, many models are not purposefully designed to conserve MSE, which can result in significant residuals. Therefore, we propose that each modeling institute provide all MSE budget terms computed consistently, as outlined in Section~\ref{subsec:before_after}.

\section{Numerically Consistent Computation vs. Physically Consistent Computation}\label{sec:accuracy}
As highlighted in Reason 1 of the previous section, the MSE budget equation is fundamentally not upheld in many discretized models, owing to inherent discrepancies between continuous and discrete calculus. This insight necessitates a reevaluation of what constitutes an \emph{accurate} calculation for the MSE budget. We distinguish two types of calculations, both deemed accurate but distinct in their approach: 1) numerically consistent computation, and 2) physically consistent computation.

To illustrate the difference between these approaches, consider a simple hypothetical example. Suppose in a certain model, we derive the following equation involving variables $A$, $B$, and $C$ from its fundamental equations:
\begin{equation}\label{eq:hypothesis}
\frac{\partial A}{\partial t} = -\nabla\cdot B\mathbf{U} + C
\end{equation}
where $C$ represents a non-dynamical source term which is irrelevant to advection. Let's also assume that while this equation can be mathematically deduced from the model's fundamental equations, it fails to hold true upon discretization, similar to the MSE budget equation. The challenge then lies in computing $-\nabla \cdot B\mathbf{U}$.

One straightforward method involves inputting the variable $B$ into the advection scheme of the model, used for transporting other tracers. This allows for computing the flux divergence in a way that is entirely consistent with the transport of other tracers. We term this approach the ``numerically consistent computation.'' However, due to the difference between the continuous calculus properties and the treatment of $B$ within the discretized advection scheme, this method is likely to fall short in completely satisfying  Eq.~\ref{eq:hypothesis}.

Alternatively, we can compute $-\nabla\cdot B\mathbf{U}$ in a different way. This method starts with an underlying assumption that the derivation of Eq.~\ref{eq:hypothesis} remains valid even in the discretized model, or that a specific advection scheme is utilized to ensure the equation's full closure. The approach then delves into the physical implications of the equation. Essentially, the equation suggests that the transport of $B$ induces a local tendency in $A$. Therefore, we interpret the physical implication of $-\nabla\cdot B\mathbf{U}$ as the change in $A$ observed before and after the advection process, calculating its value from the difference in $A$'s pre- and post-advection states. While this method may not provide a direct mathematical expression linking the tendency of $A$ with the values of $B$, it effectively captures the impact of the transport of $B$ on $A$. This methodology, predicated on the presumed validity of the derived equation in discretized models and the physical interpretation of the transport tendency, is  termed the ``physically consistent computation.''

In these contrasting approaches, the interpretation of the mathematical operator $-\nabla\cdot(\;\, \mathbf{U})$ in Eq.~\ref{eq:hypothesis} differs. In the context of numerically consistent computation, this operator is analogous to the advection scheme used for other tracers. In contrast, within the physically consistent computation, this operator might be treated as a distinct advection scheme, uniquely designed to ensure the complete closure of Eq.~\ref{eq:hypothesis}. 
This perspective is justified and aligns with modeling conventions, as it is a common practice to implement multiple differential schemes tailored to different variables within a single model.

For the computation of the column MSE flux divergence, we can apply similar approaches. In the numerically consistent computation, the MSE is first calculated, and then input into the advection scheme used for other tracers, including $\theta$ and $q_v$. This ensures that the computation of MSE transport is consistent with that of the other tracers within the model. However, this approach does not entirely resolve the issue of budget closure due to the failure of the product rule. Moreover, it leads to increased computational demands, which are typically regarded as unfavorable.

In contrast, the physically consistent computation is predicated on the assumption that the divergence operator in Eq.~\ref{eq:column_mse3} is specifically designed to ensure that the equation remains valid even after discretization. Under this assumption, the column MSE flux divergence is determined by analyzing the variations in $\langle\widetilde{h}\rangle+z_sp_s$ observed both before and after the advection process.

It is important to note that in certain models, like NICAM, which do not depend on the product rule for deriving the MSE budget equation, the numerically consistent approach should align closely with the physically consistent approach. Generally, the distinction between these two methods becomes more pronounced in models where the prognostic temperature variable is an entropic temperature quantity like potential temperature.

The choice between these two approaches hinges on specific objectives and preferences, as both methodologies offer their own advantages in terms of accuracy. In this study, we have opted for the physically consistent computation as our preferred method for the \emph{accurate} computation of the column MSE budget. This decision is driven by its simplicity and the method's broad applicability across different models with distinct structures.

\section{Computational Methods}\label{sec:method}

\subsection{Ineffective Approach: Direct Computation in the Model's Native Coordinate System}\label{subsec:direct_method}
Before delving into the specifics of the scheme for the physically consistent computation of the column MSE budget, we first address a natural assumption: that utilizing raw output data at the highest possible resolutions should tighten budget closure. Our subsequent analysis reveals that this strategy does not effectively realize the anticipated closure.

To demonstrate this, we diagnose budget closure using two alternative treatments of the transport term. In the first, we compute it in the flux-divergence form; in the second, we use the advective form. Comparing the residuals produced by each form reveals which formulation yields a smaller residual and, more importantly, exposes the limitations of an \textit{a posteriori} computation of the MSE budget.

\subsubsection{Flux-divergence Form}
\begin{figure}[!ht] 
\includegraphics[width=\textwidth]{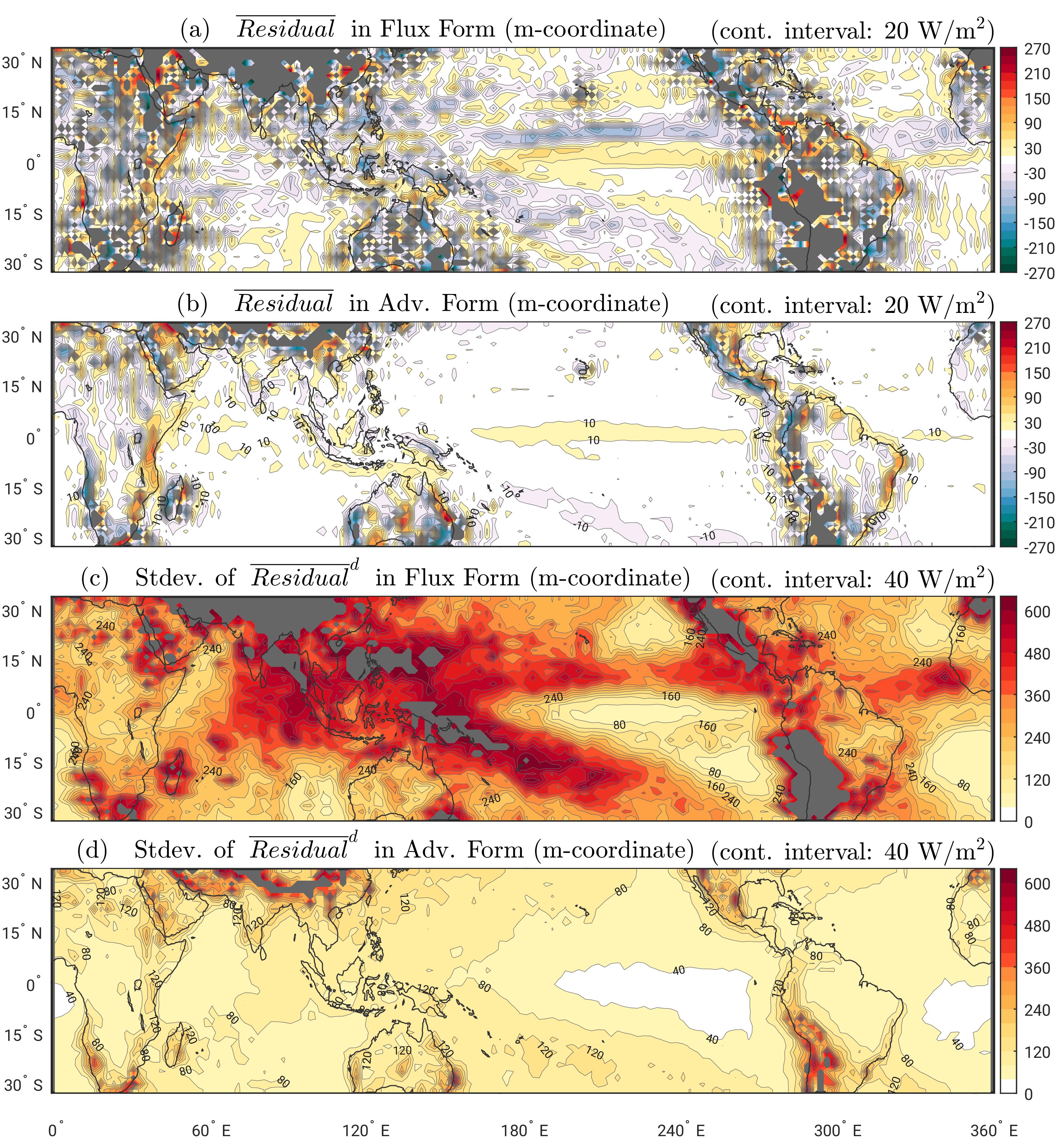} 
\caption{Annual mean and daily standard deviation of the column MSE budget residual, calculated using the raw data in the m-coordinate system, in the flux-divergence form (Eq.~\ref{eq:column_mse3}) (a, c), and in the advective form (Eq.~\ref{eq:column_mse_p}) (b, d). Contour intervals are specified in each panel, with regions of saturated contours masked in gray.}\label{fig:res_m} 
\end{figure}

In Fig.~\ref{fig:res_m}, we present the magnitude of the column MSE budget residual, computed using the raw output data in the m-coordinate system. Specifically, Figs.~\ref{fig:res_m}~(a) and (c) demonstrate the residual based on the flux-divergence form equation (Eq.~\ref{eq:column_mse3}). The term $\nabla_z\cdot\kiab{h\mathbf{v}}$ at a longitude-latitude grid point, $(i, j)$, is calculated as follows:
\begin{linenomath*}
\begin{equation}\label{eq:column_int_discrete}
\kiab{h\mathbf{v}}_{i,j} = \frac{1}{g}\sum_{k=1}^{110}h_{i,j}^{(k)}\mathbf{v}_{i,j}^{(k)}\kirb{\Delta p}_{i,j}^{(k)} \;,
\end{equation}
\end{linenomath*}
where $k$ is the index for the vertical dimension, and $\Delta p$ denotes the pressure thickness of each layer. The flux divergence is then calculated as:
\begin{linenomath*}
\begin{equation}\label{eq:flux_div_discrete}
\nabla_z\cdot\kiab{h\mathbf{v}} = \frac{1}{r\cos \varphi_j}\kirb{\frac{ \kiab{hu}_{i+\frac{1}{2},j} - \kiab{hu}_{i-\frac{1}{2},j} }{ \lambda_{i+\frac{1}{2},j}-\lambda_{i-\frac{1}{2},j} } +\frac{ \kiab{hv}_{i,j+\frac{1}{2}}\cos\varphi_{j+\frac{1}{2}}   -  \kiab{hv}_{i,j-\frac{1}{2}}\cos\varphi_{j-\frac{1}{2}} }{ \varphi_{i,j+\frac{1}{2}}-\varphi_{i,j-\frac{1}{2}} }} \;.
\end{equation}
\end{linenomath*}
Here, $r$ denotes Earth's radius, $(u,v)$ are the zonal and meridional wind components, and $\lambda$ and $\varphi$ represent longitude and latitude, respectively. The indices $i\pm1/2$ and $j\pm1/2$ correspond to the east, west, north, and south boundaries of each grid cell.  To calculate the column flux at each boundary, we first collocate all relevant variables at the respective boundary point through the spatial averaging of adjacent points. Subsequently, we perform the column integration following Eq.~\ref{eq:column_int_discrete}.

It may be surprising that the residual, calculated using the raw data and the flux-divergence form equation, is substantially greater than that computed using the postprocessed data and the advective form equation (illustrated in Figs.~\ref{fig:res_pp}(a) and (c)), both in terms of annual mean and daily standard deviation. This larger residual is due to the inherent sensitivity of the flux-divergence form to errors in mass flux divergence computations. The MSE flux divergence can be decomposed as follows:
\begin{linenomath*}
\begin{equation}
\nabla\cdot\kirb{\rho h \mathbf{U}} = \mathbf{\rho U}\cdot\nabla h + h\nabla\cdot\kirb{\rho\mathbf{U}}\;.
\end{equation}
\end{linenomath*}
Errors in the computation of $\nabla\cdot(\rho\mathbf{U})$ are exacerbated by the large values of $h$, leading to significant inaccuracies in $\nabla\cdot(\rho h\mathbf{U})$. In ModelE3, since tracers are transported using QUS, it is essential to compute mass flux divergence in alignment with this advection scheme. However, in Eq.~\ref{eq:flux_div_discrete}, the flux divergence is computed using a space-centered difference. This inconsistency results in errors in $\nabla\cdot(\rho\mathbf{U})$, which are magnified by $h$, leading to the pronounced residuals.

\subsubsection{Advective Form}
The sensitivity of the flux-divergence form to computational errors in mass flux divergence can be mitigated by employing an advective form equation. For instance, we utilize Eq.~\ref{eq:column_mse_p} to compute the residual. In this equation, the horizontal and vertical advective terms are calculated within the m-coordinate system as follows:
\begin{linenomath*}
\begin{equation}\label{eq:h_adv_m_coodinate}
\kiab{\mathbf{v}\cdot\nabla_p h} = \frac{1}{g}\int_{\sigma_t}^{\sigma_s} \mathbf{v}\cdot \kirb{\nabla_\sigma h - \frac{\partial h}{\partial p}\nabla_\sigma p   }\frac{\partial p}{\partial \sigma}\ud \sigma \;, 
\end{equation}
\end{linenomath*}
\begin{linenomath*}
\begin{equation}\label{eq:w_adv_m_coordinate}
\kiab{\omega_p\frac{\partial h}{\partial p}} =  \frac{1}{g}\int_{\sigma_t}^{\sigma_s} \omega_p \frac{\partial h}{\partial p}\frac{\partial p}{\partial \sigma}\ud \sigma \;, 
\end{equation}
\end{linenomath*}
where $\sigma$ denotes the m-coordinates, with $\sigma_s$ and $\sigma_t$ representing the values at the surface and the top of the atmosphere, respectively. $\nabla_\sigma$ is the horizontal gradient operator at constant $\sigma$. The vertical pressure velocity $\omega_p$, obtained from the model's output, is derived diagnostically through the model’s inline calculations using the mass conservation equation. The Eulerian tendency in Eq.~\ref{eq:column_mse_p} is calculated in the m-coordinate system, as defined by Eq.~\ref{eq:tend_mse_p}. For the computations in Eqs.~\ref{eq:tend_mse_p}, \ref{eq:h_adv_m_coodinate}, and \ref{eq:w_adv_m_coordinate}, we need the surface values of $h$, which are approximated by the values at the lowest model layer. Note that the choice of the $p$-coordinate system is primarily due to its consistency with the conventions of previous studies and because $\omega_p$ is included in the standard model output.

Figures~\ref{fig:res_m}(b) and \ref{fig:res_m}(d) illustrate the annual mean and daily standard deviation of the residuals, respectively, calculated from the raw data using Eq.~\ref{eq:column_mse_p}.  Compared with the corresponding panels produced with the flux–divergence formulation (Figs.~\ref{fig:res_m}(a) and \ref{fig:res_m}(c)), the advective formulation yields substantially smaller residuals in both the annual mean and the daily variability.

Each formulation has complementary strengths and weaknesses. The flux–divergence form is fully consistent with the model's native flux-form transport scheme and thus avoids reliance on $\omega_p$, whose values suffer from numerical errors---especially over regions of complex topography (Section~\ref{sec:cause}). The advective form, in contrast, eliminates the flux–divergence form's strong sensitivity to errors in the diagnosed mass-flux divergence. However, it reintroduces dependence on the error-contaminated $\omega_p$ field and remains inconsistent with the model's flux-form dynamical representation. Figure~\ref{fig:res_m} shows that the errors introduced by using the advective form are, in practice, substantially smaller than those incurred when computing flux divergence from the model output. Because $\omega_p$ is computed inline within the model, it satisfies mass conservation, making its use functionally equivalent to the mass-divergence correction schemes applied in earlier studies \cite<e.g.,>[]{trenberth_climate_1991, trenberth_using_1997, peters_analysis_2008}. We therefore conclude that the advective formulation is the more reliable choice for closing the column MSE budget.

\subsubsection{Limitations of \textit{A Posteriori} MSE-Budget Diagnostics}
Compared with Fig.~\ref{fig:res_pp}, Fig.~\ref{fig:res_m} shows discernible improvements when the column MSE budget is computed directly from the raw, highest-resolution model output. Panels~(b) and~(d) reveal that both the annual-mean and daily-variance residuals are smaller than their counterparts calculated from the postprocessed $p$-coordinate data (cf. Figs.~\ref{fig:res_pp}(a), and (c)). Over the subtropical oceans, for instance, the previously negative annual-mean residual nearly vanishes.

However, Fig.~\ref{fig:res_m} also highlights a key limitation of \textit{a posteriori} MSE-budget calculations. In the deep tropical oceans, the annual‐mean residual remains substantial, at $10$–$30$~W~m$^{-2}$. Over land, it is virtually unchanged---the magnitude and spatial pattern closely match those in Fig.~\ref{fig:res_pp}(a). The absence of improvement across continental regions indicates that adopting a terrain-following vertical coordinate does not improve budget closure there. As discussed in Section~\ref{sec:cause}, the dominant source of error is $\omega_p$ itself; because $\omega_p$ is already contaminated, refinements in vertical coordinate or grid spacing offer little leverage.

This limitation becomes even more pronounced in daily variability. Even in the raw fields, the residual standard deviation remains large---$80$–$120$~W~m$^{-2}$ over the Indian Ocean and the Western Pacific (Fig.~\ref{fig:res_m}(d))---essentially the same range seen in the fully postprocessed data in the $p$-coordinate system (Fig.~\ref{fig:res_pp}(c)).

The persistence of these residuals, regardless of computational details, suggests that they stem from irreducible errors introduced by the intertwined mechanisms outlined in Section~\ref{sec:cause}---namely Factors~1,~2,~4,~5, and~6. Because these error sources are intricately interwoven, their individual contributions cannot be disentangled by this diagnostic. Put simply, an \textit{a posteriori} computation of the column MSE budget---no matter how fine the resolution or how faithful the coordinate system---cannot achieve budget closure. Thus, the only robust remedy is to implement an inline diagnostic that tracks the MSE tendency during the simulation itself, as described in the next subsection.

\subsection{Physically Consistent Computation}\label{subsec:before_after}
\subsubsection{Process Increment Method}
\begin{figure}[!ht] 
\includegraphics[angle=0,width=14cm]{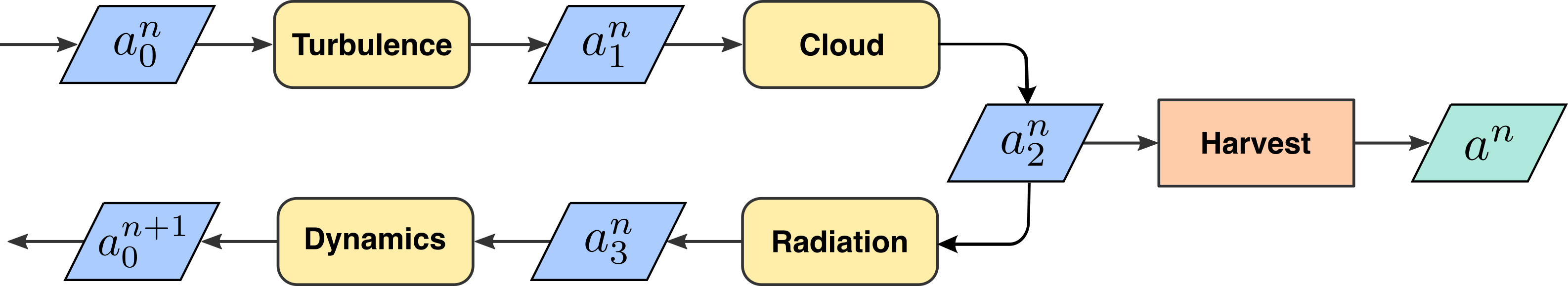} 
\caption{Flowchart illustrating the integration scheme in ModelE3. A prognostic variable, denoted as $a_0^n$ at time level $n$, undergoes forward integration to the next time level $n+1$ through a series of four distinct schemes: turbulence, cloud, radiation, and dynamics. In each step of this sequence, the scheme updates $a_0^n$ by adding its specific tendency. Notably, the variable is harvested for output, labeled as $a^n$, immediately following the completion of the cloud scheme.}\label{fig:flow} 
\end{figure}

The integration scheme of ModelE3 is structured as depicted in Fig.~\ref{fig:flow}, which illustrates the flow of the time-forward integration of an arbitrary prognostic variable $a$ at time step $n$ towards time step $n+1$ over an interval $\Delta t$. This process is segmented into four distinct schemes: turbulence, cloud, radiation, and dynamics. In each scheme, $a$ is sequentially updated by adding the tendency specific to that scheme. Consequently, by evaluating the difference in the values of $a$ before and after each scheme's application, we can determine the tendency added to $a$ by each individual scheme. This approach, which we refer to as the ``process increment method," is employed to compute the column MSE flux convergence in a physically consistent manner.

A physically consistent computation of the column MSE budget relies on the validity of Eq.~\ref{eq:column_mse3} or the existence of an advection scheme that ensures complete closure of this equation. Under that assumption, we can link each term in Eq.~\ref{eq:column_mse3} to the integration schemes shown in Fig.~\ref{fig:flow} as follows:
\begin{linenomath*}
\begin{equation}\label{eq:column_mse_label}
\frac{\partial}{\partial t}  \kirb{ \langle\widetilde{h} \rangle +z_sp_s} = \underbrace{-\nabla_z\cdot\kiab{h\mathbf{v}} + \kiab{\widetilde{\epsilon}} + \kiab{\delta}}_\text{Dynamics}  + \underbrace{L_vE + H + L_fP_{i,s} + R}_\text{Turbulence, Cloud, Radiation} \;.
\end{equation}
\end{linenomath*}

With this correspondence in place, we can calculate each term on the RHS by assessing the change in $\langle\widetilde{h} \rangle +z_sp_s$ before and after applying the corresponding scheme. Consequently, we can compute the column MSE flux convergence at time step $n$, $-(\nabla_z\cdot\kiab{h\mathbf{v}})^n$, as follows:
\begin{linenomath*}
\begin{equation}
-(\nabla_z\cdot\kiab{h\mathbf{v}})^n =  \frac{\kirb{\langle \widetilde{h}\rangle + z_sp_s}^{n+1}_0 - \kirb{\langle \widetilde{h}\rangle + z_sp_s}^n_3 }{\Delta t} -\kiab{\delta}^n  - \kiab{\widetilde{\epsilon}}^n \;,
\end{equation}
\end{linenomath*}
where $\kiab{\delta}$ is extracted from the model at each time step, and $\kiab{\widetilde{\epsilon}}$ is diagnostically computed in the m-coordinate system as:
\begin{linenomath*}
\begin{eqnarray}
\kiab{\widetilde{\epsilon}} &=&  \int_{z_s}^{z_t}\mathbf{v}\cdot\nabla_z p \ud z \\
& = & \frac{1}{g}\int_{\sigma_t}^{\sigma_s}\mathbf{v}\cdot \kirb{ \nabla_{\sigma}\phi+\frac{1}{\rho}\nabla_{\sigma} p  }\frac{\partial p}{\partial \sigma} \ud \sigma \\
&=& \frac{1}{g}\int_{\sigma_t}^{\sigma_s} \mathbf{v}\cdot\kirb{\nabla_\sigma\phi - \frac{\partial \phi}{\partial p}\nabla_\sigma p}\frac{\partial p}{\partial \sigma}\ud \sigma \;.
\end{eqnarray}
\end{linenomath*}
Here, the horizontal gradients of $p$ and $\phi$ are determined through a second-order space-centered difference. However, the specifics of computing  $\kiab{\widetilde{\epsilon}}$ may be inconsequential, given that its magnitude is considerably smaller than that of $\nabla_z\cdot\kiab{h\mathbf{v}}$. The same methodology is also applicable to the computation of the column DSE flux convergence, as elaborated in \ref{sec:dse}. Additionally, when incorporating other contributions such as the horizontal diffusion of MSE, these can also be efficiently computed using the process increment method.

Readers may recognize that the process increment method is equivalent to an indirect computation of $\nabla_z\cdot\kiab{h\mathbf{v}}$, where it is calculated as a residual derived from all other available budget terms. However, using this indirect approach requires caution: the MSE budget equation must align precisely with the physics of the analyzed model. For instance, in some models, specific and latent heats are not constant, requiring the inclusion of their effects in the computation of the Eulerian tendency. Additionally, models that account for the horizontal diffusion of MSE must include its calculation to ensure complete budget closure. Furthermore, computing the Eulerian tendency requires using instantaneous values at appropriate times of data harvesting, which should be performed within the model.

Note that the process increment approach has already been successfully applied in other modeling frameworks. For example, \citeA{chen_towards_2020} implemented a similar scheme in the Weather Research and Forecasting Model \cite<WRF;>[]{skamarock_description_2021} to compute accurate momentum and potential-temperature budgets. Likewise, \citeA{wan_condidiag10_2022} developed the online diagnostic tool CondiDiag1.0 for the Energy Exascale Earth System Model \cite<E3SM;>[]{rasch_overview_2019}, which outputs individual process tendencies (e.g., turbulence, convection, radiation) for detailed budget analyses. We therefore recommend that each modeling center adopt a comparable implementation to obtain reliable column MSE budgets, which are otherwise challenging to compute accurately.

\subsubsection{Computing Vertical MSE Advection as a Residual}\label{subsec:compute_vadv}
Finally, for comparison with previous studies, we seek to decompose the flux divergence into individual advective terms within the $p$-coordinate system. This decomposition is achieved through the following steps: First, the column-integrated horizontal MSE advection, $-\kiab{\mathbf{v}\cdot\nabla_p h}$, along with the tendency due to column mass changes, $g^{-1}h_s\partial p_s/\partial t$, are computed in the m-coordinate system. The horizontal advection is calculated following Eq.~\ref{eq:h_adv_m_coodinate}. For the column mass contribution term, $h_s$ is estimated from the value of $h$ at the lowest level, and $\partial p_s/\partial t$ is extracted directly from the model. Subsequently, the vertical advective term $\kiab{\omega_p\partial h/\partial p}$ is determined indirectly as the residual, as shown:
\begin{linenomath*}
\begin{equation}\label{eq:vadv_residual}
\kiab{ \omega_p\frac{\partial h}{\partial p}} = \nabla_z\cdot\kiab{h\mathbf{v}}+\frac{1}{g}h_s\frac{\partial p_s}{\partial t} - \kiab{\mathbf{v}\cdot\nabla_p h } \;,
\end{equation}
\end{linenomath*}
where $\nabla_z\cdot\kiab{h\mathbf{v}}$ is derived using the process increment method. 

Strictly speaking, a numerical model does not possess uniquely defined horizontal or vertical advection fields; hence any partition into advective components is artificial rather than fundamental to the model dynamics. Consequently, evaluating the accuracy or superiority of a particular partitioning scheme becomes largely a matter of interpretation.

Nevertheless, we favor this indirect approach for two main reasons. First, the horizontal wind $\mathbf{v}$ is a native prognostic variable, whereas $\omega_p$ must be reconstructed diagnostically, and therefore carries additional numerical uncertainty. Second, as shown in Section~\ref{sec:cause}, the reconstructed $\omega_p$ field exhibits clear numerical errors. We therefore estimate vertical advection using the residual approach, which is more faithful to the model's true dynamics than a direct calculation using output $\omega_p$.

\subsubsection{Caveats of the Process~Increment Method}
Although the process increment method is broadly applicable, several caveats limit its use in practice.

\noindent \textbf{Caveat 1: Consistency of the MSE budget.}
The accuracy of the method depends on a self-consistent, closed-form MSE budget equation that faithfully mirrors the model physics. However, in some cases, deriving an exact formulation may not always be possible. For example, some models may treat water substances inconsistently across the potential‐temperature equation, the ideal‐gas law, and the definitions of latent heats. These inconsistencies prevent forming a closed MSE budget from the governing equations, since the terms no longer obey thermodynamic consistency. Without a rigorously closed form---such as that in Eq.~\ref{eq:column_mse_label}---the process increment method cannot be applied.

\noindent \textbf{Caveat 2: Computational and memory demands in high‐resolution models.}
Implementing the method in convection‐permitting or global cloud‐resolving models can be resource‐prohibitive. To compute process tendencies, three‐dimensional fields, such as $T$, $z$, $q_v$, $q_i$, and $\rho$, must be stored at every substep of each physical or dynamical process.  Decomposing tendencies into vertical and horizontal advection, as required for Eq.~\ref{eq:vadv_residual}, adds further overhead: one must archive and evaluate $\mathbf{v} \cdot \nabla_{p} h$ both before and after every dynamical‐core call.  The aggregate I/O and memory footprint can be overwhelming at kilometer‐scale resolution.

Therefore, not every budget term must (or even should) be diagnosed via the process‐increment approach.  Instead, reserve the method for those tendencies that are difficult to reconstruct in offline analyses.

\section{Results and Discussion}\label{sec:results}
\subsection{Column MSE Flux Convergence}\label{subsec:results_mse_conv}

\begin{figure}[!ht] 
\includegraphics[width=\textwidth]{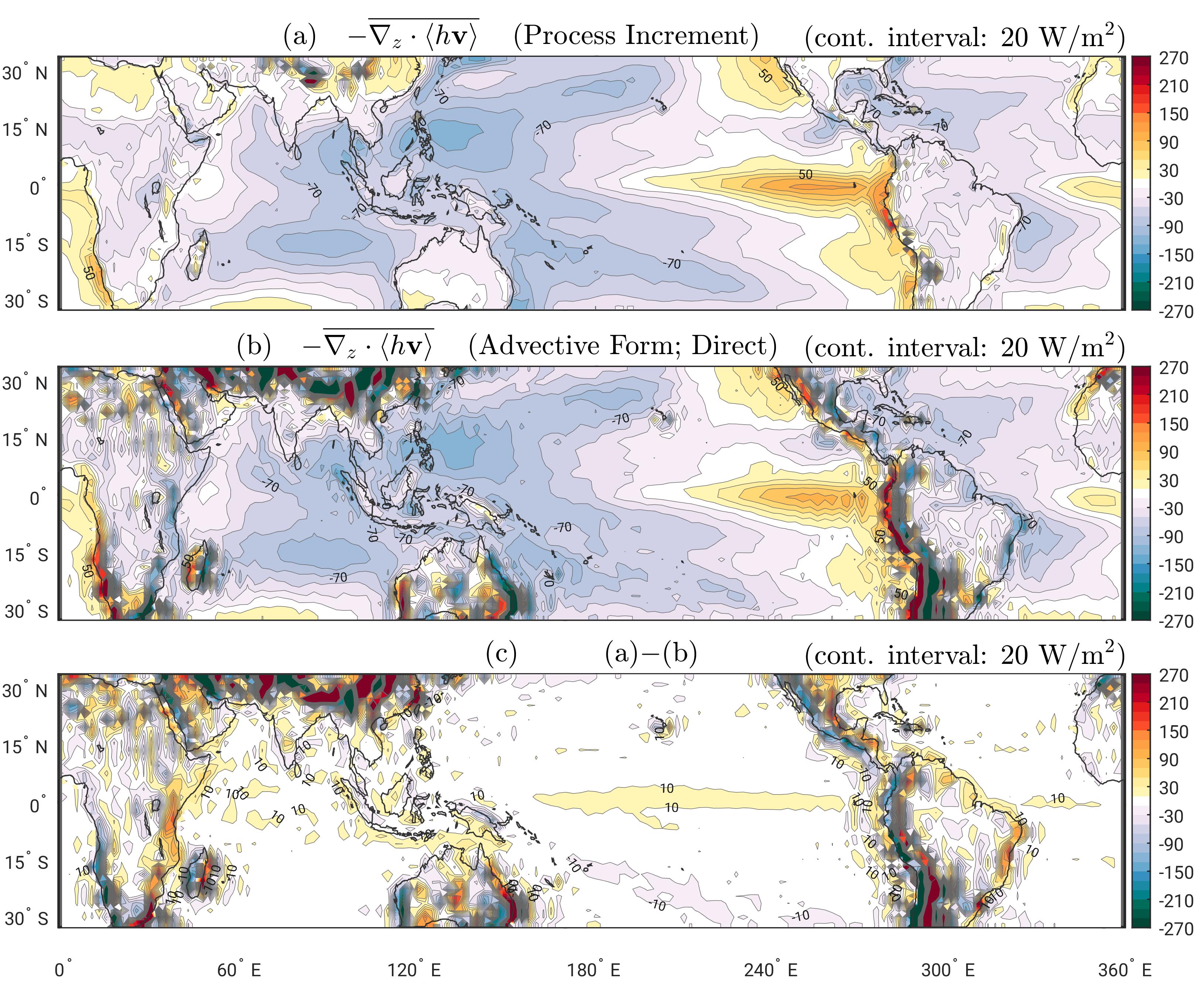} 
\caption{Annual mean of $\smash{-\nabla_z\cdot\kiab{h\mathbf{v}}}$ derived using two distinct methods and their difference: (a) the process increment method, (b) the direct computation using the raw data in the m-coordinate system at a 30-minute resolution, and (c) the difference between (a) and (b). For the direct computation in panel (b), $\smash{-\nabla_z\cdot\kiab{h\mathbf{v}}}$ is calculated in the advective form as specified in Eq.~\ref{eq:flux_advective_mse}, following the same approach as shown in Fig.~\ref{fig:res_m}(b). The contour interval for each panel is 20~W~m$^{-2}$. }\label{fig:mean_div_h} 
\end{figure}

Figure~\ref{fig:mean_div_h} presents the annual mean of $-\nabla_z\cdot\kiab{h\mathbf{v}}$ computed using two distinct methods: (a) the process increment method as detailed in Section~\ref{subsec:before_after}, and (b) the direct computation using the raw data in the m-coordinate system with the advective form equation (Eq.~\ref{eq:flux_advective_mse}). Panel (c) illustrates the difference between these two methods. Given that the budget residual in Fig.~\ref{fig:res_m}(b) originates solely from the calculation of $-\nabla_z\cdot\kiab{h\mathbf{v}}$ using the advective form, and because the process increment method is designed to ensure complete closure of the column MSE budget, the residual in Fig.~\ref{fig:res_m}(b) matches exactly the difference shown in Fig.~\ref{fig:mean_div_h}(c).

This figure highlights the limitations of using the direct computation of column MSE flux divergence (panel (b)) for accurate MSE budget analysis; the adequacy of this method varies by location and often fails to meet the quality standards necessary for precise analysis. Over the open ocean, far from coastlines, the direct computation generally reproduces patterns that agree reasonably well with the more reliable process increment method. However, over continents—especially near pronounced topography—and along coastlines, the direct computation yields spurious, noisy, wavy structures that appear unphysical. This noise arises because the model's diagnostic $\omega_p$ field is heavily contaminated by numerical errors (see Section~\ref{sec:cause}). Notably, this $\omega_p$ noise propagates widely, degrading results across all continental regions---including Africa, South America, Eurasia, and Australia---and rendering the land-based MSE budget analysis generally unreliable.

\begin{figure}[!ht]  
\includegraphics[width=\textwidth]{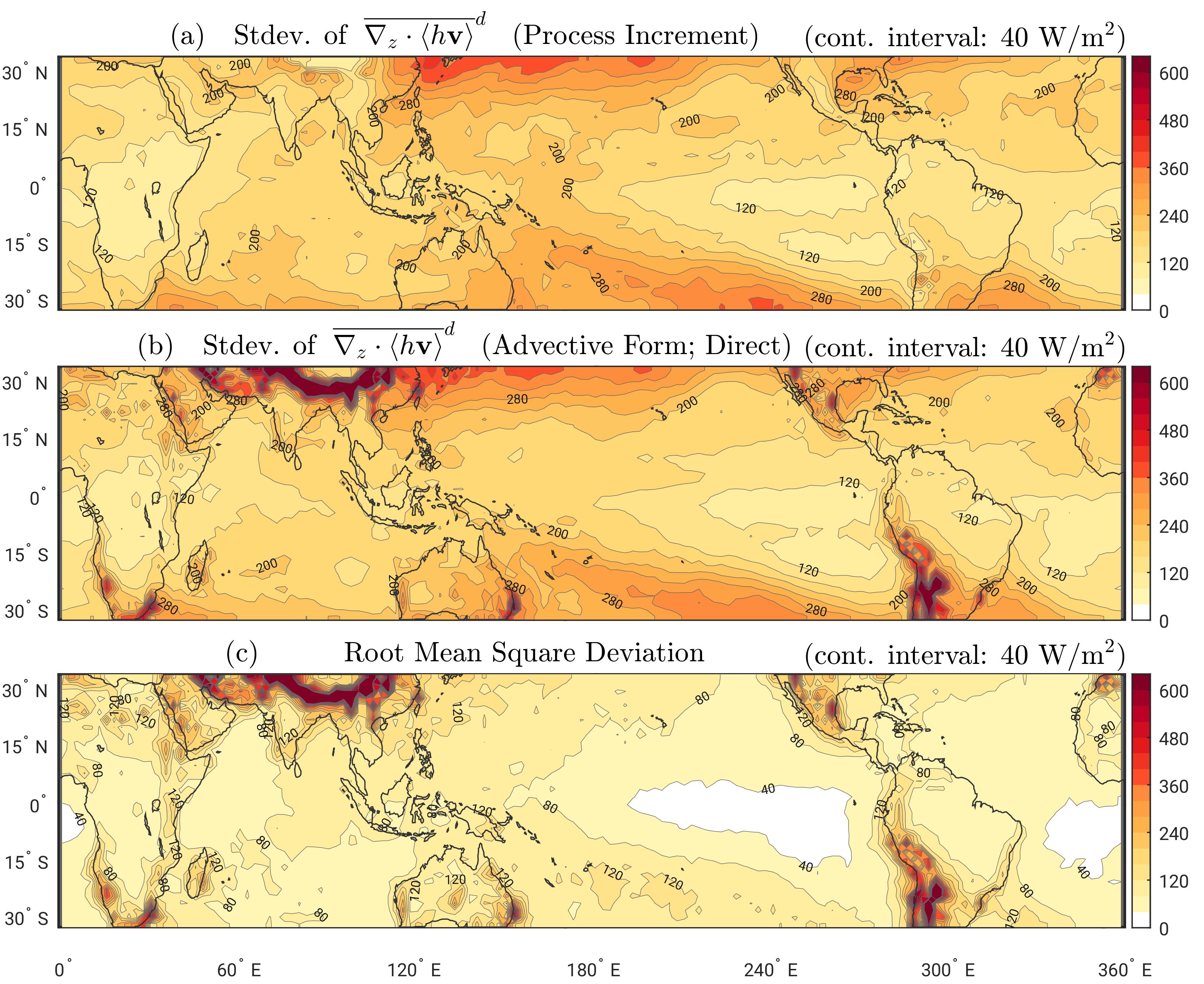} 
\caption{(a, b) Similar to Fig.~\ref{fig:mean_div_h}(a) and (b), but displaying the daily standard deviation. (c) The root mean square deviation (RMSD) between the methods used in panels (a) and (b). The contour interval for each panel is 40~W~m$^{-2}$. }\label{fig:stdev_div_h} 
\end{figure}

Figure~\ref{fig:stdev_div_h} illustrates the daily standard deviation of $\nabla_z\cdot\kiab{h\mathbf{v}}$ calculated using two distinct methods as in Fig.~\ref{fig:mean_div_h}. Panel (c) displays the root mean square deviation (RMSD) between these methods:
\begin{linenomath*}
\begin{equation}
\text{RMSD}=\sqrt{\frac{1}{N}\sum_{i=1}^N\kirb{A_i-B_i}^2} \;,
\end{equation}
\end{linenomath*}
where $A_i$ is the value computed with the process increment method at time step $i$, $B_i$ is the value computed with the direct method, and $N$ is the total length of the time series. As explained in the context of Fig.~\ref{fig:mean_div_h}(c), this RMSD should also precisely match the daily standard deviation of the residual shown in Fig.~\ref{fig:res_m}(d).

Similar to Fig.~\ref{fig:mean_div_h}, the discrepancies or RMSD between panels (a) and (b) are pronounced near topographic features and along coastlines, namely, due to numerical errors in the $\omega_p$ field.

Unlike in the annual-mean pattern shown in Fig.~\ref{fig:mean_div_h}(c), the discrepancies are also evident over the open ocean, where the $\omega_p$ field is typically less contaminated. For instance, over parts of the Western Pacific and the Indian Ocean, RMSD reaches $80$–$120$ W m$^{-2}$ (Fig.~\ref{fig:stdev_div_h}(c)), whereas the intrinsic variability of $\smash{\nabla_z\!\cdot\!\kiab{h\mathbf{v}}}$ there is $160$–$240$ W m$^{-2}$ (Fig.~\ref{fig:stdev_div_h}(a)). Thus, the residual variability is comparable to that of one of the dominant terms in the budget. As noted in Section~\ref{subsec:direct_method}, this mismatch likely arises from the tightly coupled processes summarized in Section~\ref{sec:cause}---in particular, Factors 1, 2, 4, 5, and 6---whose associated errors cannot be mitigated \textit{a posteriori}. Such persistent mismatches help to explain why MSE-budget residuals of comparable magnitude are routinely reported in studies of tropical variability \cite<e.g.,>[]{kim_propagating_2014,ren_intercomparison_2021}.

\subsection{Vertical MSE Advection}\label{subsec:results_vadv}

\begin{figure}[!ht] 
\includegraphics[width=\textwidth]{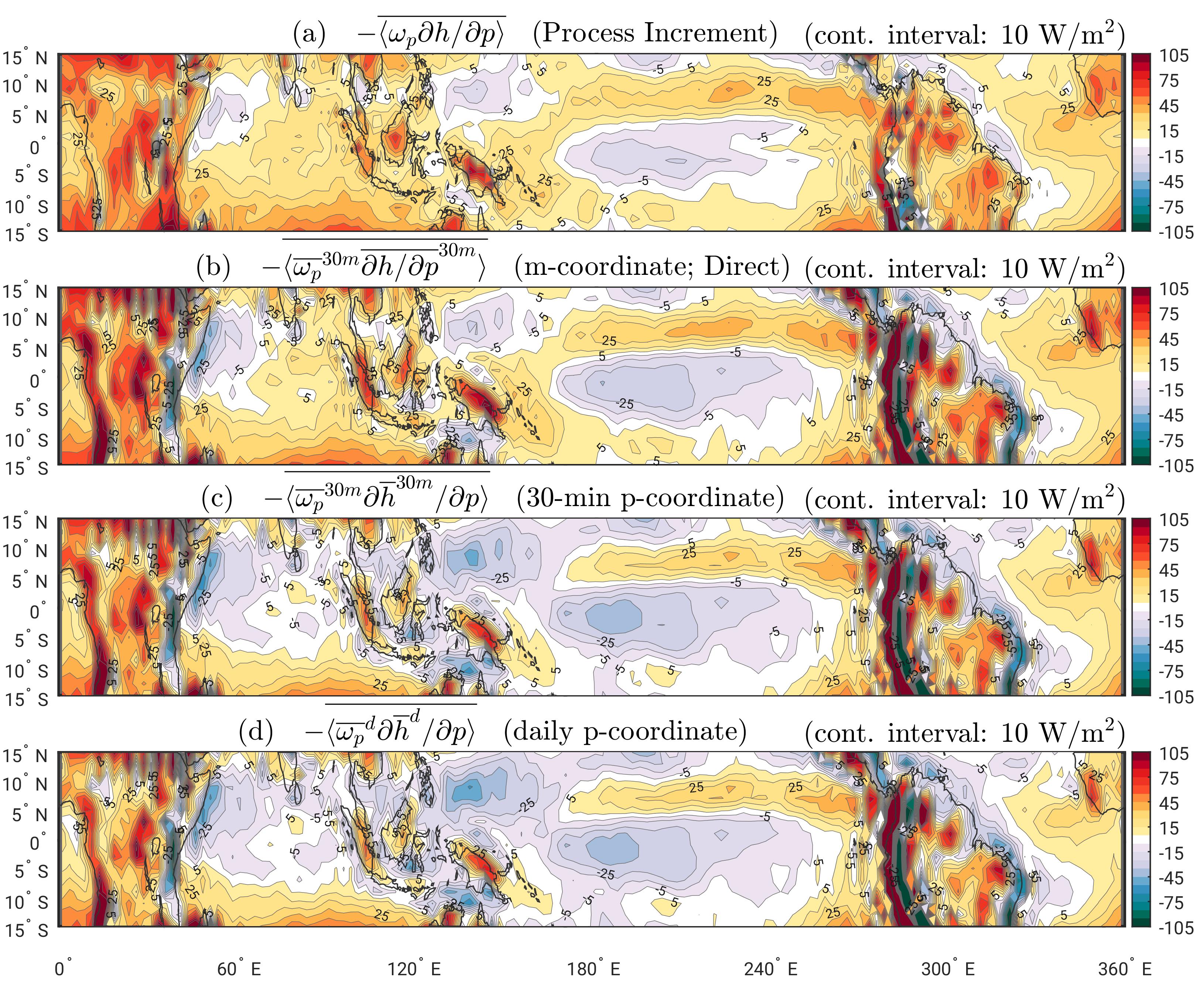} 
\caption{Annual mean of $-\kiab{\omega_p\partial h/\partial p}$ derived using four different methods: (a) the indirect computation as outlined in Section~\ref{subsec:compute_vadv}, (b) the direct computation using the raw data in the m-coordinate system at a 30-minute resolution, (c) the direct computation using the vertically interpolated data in the $p$-coordinate system at a 30-minute resolution, and (d) the direct computation using the fully postprocessed data in the daily $p$-coordinate system. Panel (d) is identical to Fig.~\ref{fig:res_pp}(b) but uses a different color scale and plot domain. The contour interval for each panel is 10~W~m$^{-2}$. }\label{fig:mean_vadv_h} 
\end{figure}

Figure~\ref{fig:mean_vadv_h} presents the annual mean of $-\kiab{\omega_p\partial h/\partial p}$ calculated using four different methods: (a) the indirect computation as outlined in Section~\ref{subsec:compute_vadv}, (b) the direct computation using the raw data in the m-coordinate system at a 30-minute resolution, (c) the direct computation using the vertically interpolated data in the $p$-coordinate system, also at a 30-minute resolution, and (d) the direct computation using the fully postprocessed data in the daily $p$-coordinate system.

Because the partitioning into individual advective components is somewhat artificial (Section~\ref{subsec:compute_vadv}), assessing the accuracy of each computed component is inherently challenging. Nevertheless, the indirect approach shown in panel~(a)---which relies solely on the model's native prognostic variables and thus avoids the potentially erroneous $\omega_p$ fields---should provide a depiction that more closely reflects the model's true dynamics than the direct method depicted in panel~(b). Moreover, any features common to panels~(a) and~(b) are largely insensitive to the computational pathway and can therefore be considered robust.

Consistent with the results in Fig.~\ref{fig:mean_div_h}, the direct estimates (Figs.~\ref{fig:mean_vadv_h}(b)--(d)) are contaminated by wavelike noise over land and along coastlines. Because the error resides in the $\omega_p$ field itself, its spatial pattern and amplitude are nearly identical across panels (b), (c), and (d) despite the different coordinate systems and temporal averaging employed. By circumventing the problematic $\omega_p$ field, the indirect calculation in panel~(a) avoids these artifacts altogether.

The figure also highlights the dependence of vertical-advection estimates on the vertical coordinate. Transitioning from the raw m‑coordinate to the $p$‑coordinate (panel (b) $\rightarrow$ (c)) can introduce errors large enough to reverse the sign of $-\kiab{\omega_p\partial h/\partial p}$, notably over the oceanic ITCZ. In contrast, reducing the temporal resolution from 30‑min to daily averages (panel (c) $\rightarrow$ (d)) leaves the mean field largely unaffected. Thus, vertical coordinate choice---rather than temporal averaging---is the dominant source of uncertainty in the mean vertical‑advection estimate; this sensitivity may warrant a re‑examination of conventional views of the ITCZ energy budget.

Previous research suggests that in the warm-pool regions including the Western Pacific and the Indian Ocean, $\omega_p$ profiles are top-heavy, leading to net MSE export by vertical circulations (i.e., $-\overline{\kiab{\omega_p\partial h/\partial p}} < 0$), whereas in the Eastern Pacific, bottom‑heavy profiles import MSE (i.e., $-\overline{\kiab{\omega_p\partial h/\partial p}} > 0$) \cite<e.g.,>[]{back_geographic_2006, raymond_mechanics_2009, inoue_evidence_2021}. This behavior is often interpreted in terms of gross moist stability (GMS): positive GMS implies MSE export, and negative GMS implies import \cite<e.g.,>[]{neelin_modeling_1987, Neelin_2008_10, raymond_mechanics_2009, inoue_gross_2015}. Generally, the GMS averaged across the entire ITCZ is presumed to be positive, suggesting that vertical circulations serve as a mechanism for exporting energy from the tropics to the subtropics. This view of positive GMS has underpinned many theoretical works, including theories on the MJO \cite<e.g.,>[]{sobel_weak_2001, fuchs_simple_2007, raymond_convectively_2007, sugiyama_moisture_2009a, sobel_idealized_2012, adames_mjo_2016}, the dynamics of the ITCZ/Hadley circulation \cite<e.g.,>[]{neelin_modeling_1987, Neelin_2008_10, byrne_energetic_2016, ahmed_process_2023}, and the Walker circulation \cite<e.g.,>[]{bretherton_simple_2002}. However, Fig.~\ref{fig:mean_vadv_h} raises questions about this traditional understanding, suggesting that the typical characterization of vertical MSE advection as an energy-exporting mechanism in the tropics may warrants further investigation.

The conventional perspective is supported by Fig.~\ref{fig:mean_vadv_h}(c), which uses the vertically interpolated data and exhibits negative values of $-\overline{\kiab{\omega_p\partial h/\partial p}}$ over the Western Pacific, parts of the Indian Ocean, and the South Pacific Convergence Zone (SPCZ). This export signal over the warm-pool regions becomes slightly more pronounced in panel~(d), which is derived using the fully postprocessed data in the daily $p$-coordinate system. However, this view is challenged by both the indirect calculation (panel~(a)) and the direct calculation using the raw data (panel~(b)). In fact, except for a small region in the Western Pacific, $-\overline{\kiab{\omega_p\partial h/\partial p}}$ is positive across the entire oceanic ITCZ, indicating that vertical advection is actually importing energy from the subtropics into the tropics. This significant discrepancy is attributed to the vertical interpolation from the model's 110 native vertical levels to 32 pressure levels. This interpolation tends to decrease the value of $-\overline{\kiab{\omega_p\partial h/\partial p}}$, thereby shifting the field toward a greater apparent export (i.e., a more pronounced blue color in Fig.~\ref{fig:mean_vadv_h}), particularly over the warm-pool regions. This systematic error leads to an overestimation of MSE export within the warm-pool regions when analyzed within the $p$-coordinate system, and daily averaging slightly exacerbates this overestimation. More detailed analyses, including geographic variations of the profiles of $\omega_p$ and MSE, as well as the decomposition into contributions from mean circulations and transients, will be presented in Part~2 of this two-part series.

It must be emphasized that we cannot conclusively determine that the tropical GMS is negative from this study alone. ModelE3 may simply overestimate MSE import relative to other models or observations. Instead, the most significant implication of our findings is that changes in the vertical coordinates used for analysis can be sufficient to reverse the sign of GMS. Given the critical role of vertical MSE advection and GMS in tropical climate dynamics, we recommend that each modeling institute provide vertical MSE advection data both indirectly computed (as in panel (a)) and directly computed using raw data (panel (b)). This would enable model users to thoroughly analyze vertical MSE advection and GMS.

\section{Summary and Conclusions}\label{sec:summary}
The primary limitation in analyzing the column MSE budget lies in its computational challenges, particularly in \textit{a posteriori} calculations of the transport term. Typically, significant residuals are observed in the column MSE budget computations, with magnitudes comparable to other major budget terms both in annual mean and daily variability. Such large residuals complicate drawing clear conclusions about the roles of individual physical processes in affecting the column MSE. In this study, we address this challenge by implementing an inline diagnostic code for accurately computing the column MSE budget using GISS ModelE3. We demonstrate how the implemented methodology enhances the accuracy of column MSE budget analysis and argue that such precise computation is essential for reliable MSE budget analysis.

To ensure an accurate computation of the column MSE budget, it is essential to derive the conservation law that aligns precisely with the model’s physics. We conducted a rigorous derivation of the column MSE budget equation, detailed in Section~\ref{sec:derivation}. Typically, the column MSE budget equation is approximated by omitting 1) the work done against the pressure-gradient force, 2) the pressure tendency, 3) surface snowfall, and 4) frictional dissipation (Eq.~\ref{eq:column_fmse_approx}). This approximation generally does not impact the budget residual (Fig.~S1), as the residual is more significantly influenced by other fundamental factors which are identified in this study.

In Section~\ref{sec:cause}, we identified seven reasons for the failure to close the MSE budget in ModelE3:
\begin{enumerate}
\item Failure of the product rule: In a discretized model, certain continuous-calculus identities---most notably the product rule---no longer hold, rendering the algebraic conversion from the potential-temperature equation to its DSE form invalid.
\item Impact of mass filtering on MSE transport: Numerical filters applied to the mass field alter the column mass, which affects temperature through adiabatic heating or cooling due to mass changes. This process, a part of vertical DSE advection, is generally impossible to reconstruct from standard model outputs.
\item Challenges in reconstructing flux divergence using standard outputs: Accurately reconstructing flux divergence from model outputs is generally unattainable, leading to significant errors.
\item Inconsistency of advective form with the model's dynamical core: Because the dynamical core of ModelE3 solves the flux-form equations, evaluating MSE transport in advective form inevitably introduces a systematic residual.
\item Numerical errors in output $\omega_p$: The vertical wind $\omega_p$, derived through ModelE3's diagnostic computations, is contaminated by numerical errors especially over continents and near coastlines.
\item Timing of model output harvesting: Because diagnostic fields are sampled at discrete times, they may not coincide with the precise instants at which the model advects those quantities, causing inconsistencies.
\item Vertical interpolation and temporal averaging: Postprocessing operations—particularly vertical interpolation and temporal averaging—introduce further errors into the MSE budget.
\end{enumerate}
These factors interact in a highly complex manner, reinforcing one another and rendering any \textit{a posteriori} reconstruction of the MSE transport term infeasible, irrespective of model resolution or the sophistication of the postprocessing applied to standard outputs.

One important supplementary implication from this study is the potential advantages of using an advective form equation (Eqs.~\ref{eq:column_mse_z} or \ref{eq:column_mse_p}) over a flux-divergence form (Eq.~\ref{eq:column_mse3}) for computing the column MSE budget. Although the advective form does not align with the model's actual transport processes and relies on potentially contaminated $\omega_p$, the errors resulting from the advective form are considerably smaller than those from inaccuracies in reconstructing flux divergence from model outputs.

In models where the prognostic temperature variable is an entropic quantity, the MSE budget does not close if MSE is treated in the same manner as other conserved tracers. This fact necessitates a reevaluation of what constitutes an \emph{accurate} computation of the MSE budget. We distinguish between two types of computations: 1) numerically consistent computation, and 2) physically consistent computation. Both approaches can be considered accurate, but only the physically consistent approach enables proper closure of the budget.

In the numerically consistent computation, MSE is initially calculated and then input into the advection scheme used for other conserved tracers. Although this approach ensures that the computation of MSE transport aligns with that of other tracers within the model, it does not achieve complete closure of the MSE budget. Moreover, this approach increases computational demands, which is generally undesirable.

Our preferred approach, the physically consistent computation, is predicated on the assumption that the MSE budget equation (Eqs.~\ref{eq:column_mse}, \ref{eq:column_mse2}, or \ref{eq:column_mse3}) ought to remain valid in discretized models, or equivalently, that there exists an advection scheme ensuring its complete closure. Based on this assumption, we compute the dynamical tendency of column MSE by using the process increment method, which calculates the difference in the values of $\langle\widetilde{h} \rangle +z_sp_s$ before and after the dynamical integration scheme is applied. Additionally, we calculate the column-integrated vertical MSE advection indirectly as a residual, given the column-integrated horizontal advection, the tendency from column mass changes, and the column MSE flux divergence computed using the process increment method. This approach effectively bypasses the use of erroneous $\omega_p$. The process increment approach has already been successfully implemented in other modeling frameworks \cite<e.g.,>[]{chen_towards_2020, wan_condidiag10_2022}, indicating potential applicability across various models. Furthermore, it can be used to compute the column DSE flux divergence, as elaborated in \ref{sec:dse}.

The process increment method for calculating column MSE flux divergence and the indirect computation of vertical MSE advection enable precise MSE budget analysis over all regions, including continental areas, as shown in Figs.~\ref{fig:mean_div_h}, \ref{fig:stdev_div_h}, and \ref{fig:mean_vadv_h}. This level of precision is unattainable through direct computations using standard model outputs, regardless of the meticulousness of the computational details.

Most critically, Fig.~\ref{fig:mean_vadv_h} highlights the significant impact of vertical interpolation into the $p$-coordinate system, showing that such interpolation can introduce errors substantial enough to reverse the sign of $-\overline{\kiab{\omega_p\partial h/\partial p}}$ over the oceanic ITCZ. This underscores the sensitivity of vertical MSE advection computations to vertical coordinate systems, necessitating careful consideration in MSE budget analysis.

In the $p$-coordinate system, our results show negative $-\overline{\kiab{\omega_p\partial h/\partial p}}$ over the warm-pool regions, consistent with previous studies that suggest top-heavy $\omega_p$ profiles in these regions lead to MSE export, while bottom-heavy profiles in the Eastern Pacific lead to MSE import. This view, encapsulated by the concept of GMS, has been foundational in many theories of tropical climate dynamics. However, when using the model’s native vertical coordinates, the sign of $-\overline{\kiab{\omega_p\partial h/\partial p}}$ can be flipped, particularly over the warm-pool regions, indicating a potential energy import via vertical circulations from the subtropics into the tropics. These findings suggest that vertical interpolation errors may be significant enough to alter the computed GMS. Therefore, the conventional beliefs about vertical MSE advection and GMS may require reevaluation in light of these sensitivities. 

Because the present conclusions are drawn from a single model, their robustness must be tested.  Future work should repeat the analysis with additional models, again employing the process increment method for computing the transport terms and explicitly quantifying the sensitivity of vertical MSE advection to the vertical-coordinate choice.

The column MSE budget underpins our understanding of phenomena such as the MJO, tropical-cyclone genesis, convective aggregation, monsoon dynamics, and their responses to climate change. Progress in each area is hindered when the budget cannot be closed. We therefore recommend that each modeling institute provide all necessary terms to comprehensively close the column MSE budget. Making these terms readily available will enable more reliable and thorough MSE budget analysis and, ultimately, more robust climate projections.

This paper establishes the foundations for the upcoming second part of our two-part series. Using the column MSE budget terms accurately derived as outlined in this paper, the next installment provides a comprehensive assessment of each term, including those that are often neglected in conventional MSE budget approximations, examining their significance and broader implications. Additionally, we evaluate the impact of postprocessing procedures on MSE budget analysis via a detailed comparison of budget terms derived through our new methodology against those derived from postprocessed data. The impact on vertical MSE advection is thoroughly examined, highlighting how accurate MSE budget computations can significantly affect the understanding of tropical climate dynamics.  Through this rigorous analysis, we aspire to provide deeper insights and enhance understanding in the field of MSE budget computations.

\newpage
\appendix

\section{Derivation of the Total Energy Budget Equation}\label{subsec:total}

The derivation begins with the three-dimensional momentum equation\footnote{Some authors introduce a negative sign in the viscous term of Eq.~\ref{eq:momentum} \cite<e.g.,>[]{peixoto_physics_1992, randall_introduction_2015}, while others do not \cite<e.g.,>[]{kundu_fluid_2008, kuo_introduction_2012}. In the first convention, the viscous stress tensor is defined to be negatively proportional to the strain rate, canceling the negative sign. In this study, we adopt the latter convention.}:
\begin{linenomath*}
\begin{equation}\label{eq:momentum}
\frac{D\mathbf{U}}{Dt} = -2\mathbf{\Omega}\times\mathbf{U}-\nabla\phi - \frac{1}{\rho} \nabla p + \frac{1}{\rho}\nabla\cdot\kitens{\tau} \;,
\end{equation}
\end{linenomath*}
where $\mathbf{\Omega}$ represents the Earth's angular velocity vector, and the symbol $\times$ denotes the cross product. The term $\kitens{\tau}$ denotes the second-order viscous stress tensor, making $\smash{\nabla\cdot\kitens{\tau}}$ a vector. Detailed descriptions of the stress tensor and the tensor operations are provided in \ref{sec:tensor}.

By applying $\mathbf{U}\cdot$ to Eq.~\ref{eq:momentum}, we derive the kinetic energy equation:
\begin{linenomath*}
\begin{equation}\label{eq:kinetic_energy}
\frac{DK}{Dt} = -\widetilde{\epsilon} + \frac{1}{\rho}\mathbf{U}\cdot\kirb{\nabla\cdot\kitens{\tau}}\;,
\end{equation}
\end{linenomath*}
where $K \equiv (\mathbf{U}\cdot \mathbf{U})/2$ represents the kinetic energy per unit mass. This equation suggests that in situations where the viscous force is minimal, such as in the free troposphere, an accelerating parcel (i.e., $DK/Dt > 0$) results in negative $\widetilde{\epsilon}$, thereby acting as a net sink of enthalpy or MSE, as indicated in Eq.~\ref{eq:fmse}. In simpler terms, a parcel that is speeding up tends to become cooler because of adiabatic expansion.

By applying the tensor identity from Eq.~\ref{eq:tens_id} and introducing the double contraction operator (Eq.~\ref{eq:double_contraction}), we can decompose $\rho^{-1}\mathbf{U}\cdot(\nabla\cdot\kitens{\tau})$ into two distinct terms:
\begin{linenomath*}
\begin{equation}\label{eq:tau_decompose}
\frac{1}{\rho}\mathbf{U}\cdot\kirb{\nabla\cdot\kitens{\tau}} = \frac{1}{\rho}\nabla\cdot\kirb{\kitens{\tau}\cdot\mathbf{U}}-\delta \;,
\end{equation}
\end{linenomath*}
where
\begin{linenomath*}
\begin{equation}
\delta \equiv \frac{1}{\rho}\kitens{\tau}:\nabla\mathbf{U}\;.
\end{equation}
\end{linenomath*}
The term $\rho^{-1} \nabla \cdot \left( \kitens{\tau}\cdot \mathbf{U} \right)$ represents the rate of total work done by the viscous force per unit mass. Meanwhile, $\delta$ signifies the rate of viscous dissipation of kinetic energy per unit mass. For a Newtonian fluid, $\delta$ is positive-definite and thus acts as a net sink of kinetic energy, as indicated by the negative sign in Eq.~\ref{eq:tau_decompose} [e.g., refer to Ch.13 of \citeA{kundu_fluid_2008}]. This term illustrates the transformation of kinetic energy into internal energy and is reflected in the potential temperature equation (Eq.~\ref{eq:potential_temp}) as frictional heating.

By incorporating Eq.~\ref{eq:kinetic_energy} into the MSE budget equation (Eq.~\ref{eq:fmse}), and utilizing Eq.~\ref{eq:tau_decompose}, we derive the total energy budget equation:
\begin{linenomath*}
\begin{equation}\label{eq:total_energy}
\frac{D}{Dt}\kirb{h+K} = \frac{1}{\rho}\frac{\partial p}{\partial t}- \frac{1}{\rho}\nabla\cdot \kirb{\mathbf{R} + L_f\mathbf{P}_i  +\mathbf{F}_h -\kitens{\tau}\cdot\mathbf{U}}\;.
\end{equation}
\end{linenomath*}

By multiplying Eq.~\ref{eq:total_energy} by $\rho$ and incorporating the mass conservation (Eq.~\ref{eq:mass_conservation}) and the ideal gas law (Eq.~\ref{eq:ideal_gas}), we derive the flux-divergence form of the total energy budget equation:
\begin{linenomath*}
\begin{equation}\label{eq:flux_div_total_energy}
\frac{\partial}{\partial t}\kicb{\rho\kirb{h+K}} = -\nabla\cdot\kicb{\rho\kirb{h+K}\mathbf{U}} + \frac{\partial}{\partial t}\kirb{\rho R_dT}- \nabla\cdot \kirb{\mathbf{R} + L_f\mathbf{P}_i  +\mathbf{F}_h -\kitens{\tau}\cdot\mathbf{U}} \;.
\end{equation}
\end{linenomath*}
Following a similar procedure to that in \ref{sec:from_flux_column}, integrating Eq.~\ref{eq:flux_div_total_energy} with respect to $z$ yields the column-integrated total energy budget equation:
\begin{linenomath*}
\begin{eqnarray}\label{eq:column_total_energy}
\frac{\partial}{\partial t} \kiab{h-R_dT+K} = &-&\nabla_z\cdot\kiab{\kirb{h+K}\mathbf{v}} +R +L_fP_{i,s}+L_vE+H \\ \nonumber
&+& \kiab{\frac{1}{\rho}\nabla_z\cdot\kirb{\kitens{\tau}\cdot\mathbf{U}}_h}- \kirb{\kitens{\tau}\cdot\mathbf{U}}_s  \;.
\end{eqnarray}
\end{linenomath*}
Here, as in Eq.~\ref{eq:column_mse}, we assume the horizontal components of $\mathbf{R}$, $\mathbf{P}_i$, and $\mathbf{F}_h$ to be zero, consistent with the assumptions in ModelE3. However, the horizontal component of $\smash{\kitens{\tau}\cdot\mathbf{U}}$, denoted as $\kirb{\kitens{\tau}\cdot\mathbf{U}}_h$, typically has non-zero values due to the parameterization of momentum damping\footnote{Dynamical cores commonly apply horizontal momentum damping—implemented as Laplacian, hyperdiffusive, or Smagorinsky-type viscosities—to maintain numerical stability. As a result, the horizontal viscous flux of kinetic energy is generally nonzero. Although this term is typically small and may be negligible for many energy-budget diagnostics, we retain it to enforce exact budget closure, which is essential for the ``process increment method'' employed in this study.}, necessitating its vertical integration term. The term $\kirb{\kitens{\tau}\cdot\mathbf{U}}_s$ represents the surface value of the vertical component of $\kitens{\tau}\cdot \mathbf{U}$, which is assumed to be zero at the top of the atmosphere but non-zero at the surface.

For the computation of the column energy budget, both the column MSE budget equations (Eqs.~\ref{eq:column_mse} and \ref{eq:column_mse2}) and the column total energy budget equation (Eq.~\ref{eq:column_total_energy}) present feasible approaches. However, we find that the column MSE budget equations align more effectively with the objectives of our study. The primary reason for this preference is the complexity in obtaining the last two terms of Eq.~\ref{eq:column_total_energy} directly from a model. Therefore, for precise and practical computation of the energy budget, we utilize the column MSE budget equation in this study.

\section{Stress Tensor and Tensor Algebra}\label{sec:tensor}
The stress on a fluid particle, defined as force per unit area, requires nine components for complete specification and is represented as a second-order tensor. For example, the viscous stress tensor in Eq.~\ref{eq:momentum} is expressed as:
\begin{linenomath*}
\begin{equation}
\kitens{\tau} \equiv
\begin{bmatrix}
\tau_{11} & \tau_{12} & \tau_{13} \\
\tau_{21} & \tau_{22} & \tau_{23} \\
\tau_{31} & \tau_{32} & \tau_{33}
\end{bmatrix} \;,
\end{equation}
\end{linenomath*}
where $\tau_{ij}$ represents the $j$-th component of the force acting on a surface, with its outward normal vector pointing in the $i$-direction. It is essential to note that this stress tensor is symmetric (i.e., $\tau_{ij}=\tau_{ji}$) to avoid the scenario of infinite angular acceleration in an infinitesimal fluid particle.

In this study, we employ four distinct tensor operations to manipulate the wind vector and the viscous stress tensor. These operations are the contraction of tensors ($\cdot$), the double contraction of tensors ($:$), the divergence of a tensor ($\nabla\cdot$), and the gradient of a vector ($\nabla$). We provide explicit definitions for each operation, as different authors may use varying conventions.

First, we define the contraction of two tensors of arbitrary order, $\kitens{A}$ and $\kitens{B}$, as follows:
\begin{linenomath*}
\begin{equation}
\kitens{A}\cdot\kitens{B} = \sum_k A_{\ldots k}B_{k\ldots}\;.
\end{equation}
\end{linenomath*}
In this operation, the last index of the first tensor ($\kitens{A}$) is equated to the first index of the second tensor ($\kitens{B}$), and a summation is performed over this shared index. When applied to two vectors (or first-order tensors with one index), this operation simplifies to $\sum A_kB_k$, which represents the dot product of the vectors. Notably, in Einstein notation, the summation symbol is often omitted.

Second, utilizing this index convention, we define the divergence of the viscous stress tensor in Cartesian coordinates. The $j$-component of this divergence is expressed as:
\begin{linenomath*}
\begin{equation}
\left( \nabla\cdot\kitens{\tau} \right)_j = \sum_{i=1}^3\nabla_i\tau_{ij}=\sum_{i=1}^3\frac{\partial \tau_{ij}}{\partial x_i} \;.
\end{equation}
\end{linenomath*}
Some authors define this as $\partial\tau_{ij}/\partial x_j$, but both formulations are equivalent due to the symmetry of the stress tensor, where $\tau_{ij}=\tau_{ji}$.

Third, similarly to the contraction, we define the double contraction of two tensors of arbitrary order as follows:
\begin{linenomath*}
\begin{equation}\label{eq:double_contraction}
\kitens{A}:\kitens{B} = \sum_j\sum_k A_{\ldots jk}B_{jk\ldots}\;.
\end{equation}
\end{linenomath*}
When applied to two second-order tensors (or matrices), this becomes $\sum\sum A_{ij}B_{ij}$, which is a scalar and called the Frobenius inner product.

Finally, we define the gradient of a vector $\mathbf{U}$ in Cartesian coordinates, whose $ij$-component is expressed as:
\begin{linenomath*}
\begin{equation}
\left( \nabla \mathbf{U} \right)_{ij} = \frac{\partial U_j}{\partial x_i}\;.
\end{equation}
\end{linenomath*}

With the aforementioned definitions, we can derive the following tensor identity:
\begin{linenomath*}
\begin{equation}\label{eq:tens_id}
\mathbf{U}\cdot \left( \nabla\cdot\kitens{\tau} \right) = \nabla \cdot \left( \kitens{\tau}\cdot \mathbf{U}   \right) -\kitens{\tau}:\nabla\mathbf{U}\;.
\end{equation}
\end{linenomath*}
This can be proven as follows:
\begin{linenomath*}
\begin{eqnarray}
\mathbf{U}\cdot \left( \nabla\cdot\kitens{\tau} \right) & =& \sum_j U_j \left( \nabla\cdot\kitens{\tau}  \right)_j \\
& =& \sum_j  U_j \sum_i \frac{\partial \tau_{ij}}{\partial x_i}  \\
& =& \sum_i\sum_j \left\{ \frac{\partial (\tau_{ij} U_j) }{\partial x_i } -\tau_{ij}\frac{\partial U_j}{\partial x_i}   \right\} \\
& =& \sum_i\frac{\partial}{\partial x_i}  \sum_j \tau_{ij}U_j  - \sum_i\sum_j \tau_{ij}\frac{\partial U_j}{\partial x_i} \\
& =& \sum_i \frac{\partial}{\partial x_i} \left( \Bar{\Bar{\tau}}\cdot \mathbf{U}  \right)_i - \sum_i\sum_j\tau_{ij} \left( \nabla \mathbf{U} \right)_{ij} \\
& =& \nabla \cdot \left( \kitens{\tau}\cdot \mathbf{U}   \right) -\kitens{\tau}:\nabla\mathbf{U}\;.
\end{eqnarray}
\end{linenomath*}

\section{Column Integration of Flux-Divergence Form}\label{sec:from_flux_column}
We obtain Eq.~\ref{eq:column_mse} from Eq.~\ref{eq:flux_div_mse} as follows:
\begin{linenomath*}
\begin{eqnarray}
\int_{z_s}^{z_t}\kicb{ \frac{\partial}{\partial t}\kirb{\rho h} + \nabla\cdot\kirb{\rho h\mathbf{U}} } \ud z& = &  \int_{z_s}^{z_t}\kicb{ \frac{\partial}{\partial t}\kirb{\rho h} + \nabla_z\cdot\kirb{\rho h\mathbf{v}} + \frac{\partial}{\partial z} \kirb{\rho h w}} \ud z \label{eq:flux_div1}\\
& = & \frac{\partial}{\partial t}\int_{z_s}^{z_t}\rho h \ud z + \nabla_z\cdot\int_{z_s}^{z_t}\rho h\mathbf{v}\ud z \nonumber \label{eq:flux_div2}\\ 
&+& \rho_sh_s\kirb{\frac{\partial z_s}{\partial t} + \mathbf{v}_s\cdot\nabla_z z_s -w_s}\;, 
\end{eqnarray}
\end{linenomath*}
where the subscript $s$ represents a surface values. From Eq.~\ref{eq:flux_div1} to Eq.~\ref{eq:flux_div2}, we apply the Leibniz integral rule with the assumption that $\rho=0$ at the top of the atmosphere. Since we have $w_s\equiv Dz_s/Dt = \partial z_s/\partial t + \mathbf{v}_s\cdot\nabla_z z_s$ (generally, $\partial z_s/\partial t =0$), the last term of Eq.~\ref{eq:flux_div2} vanishes. Consequently, we obtain
\begin{linenomath*}
\begin{equation}
\int_{z_s}^{z_t}\kicb{ \frac{\partial}{\partial t}\kirb{\rho h} + \nabla\cdot\kirb{\rho h\mathbf{U}} } \ud z = \frac{\partial \kiab{h}}{\partial t} + \nabla_z\cdot\kiab{h\mathbf{v}}\;.
\end{equation}
\end{linenomath*}

\section{The Sub-grid Contribution of $\widetilde{\epsilon}$}\label{subsec:hydrostatic}
Here, we clarify why the sub-grid contribution of $\widetilde{\epsilon}$ can be absorbed into $\nabla \cdot \mathbf{F}_t$ and $\delta$.  Let $\kisb{\;\;}$ denote a spatial (grid-box) average.  Taking the average of $\rho\widetilde{\epsilon}$ separates resolved and unresolved parts:
\begin{linenomath*}
\begin{equation}\label{eq:mse_ave_and_eddy}
\kisb{\rho \widetilde{\epsilon}}=\rho \widetilde{\epsilon}_{ave} + \rho \widetilde{\epsilon}_{eddy} \;,
\end{equation}
\end{linenomath*}
where
\begin{linenomath*}
\begin{equation}\label{eq:pressure_gradient_work}
\rho \widetilde{\epsilon}_{ave} \equiv \kisb{\mathbf{v}}\cdot\nabla_z\kisb{p} + \kisb{w}\kirb{\frac{\partial \kisb{p}}{\partial z} + \kisb{\rho}g  } \;,
\end{equation}
\end{linenomath*}
and
\begin{linenomath*}
\begin{align}
\rho\widetilde{\epsilon}_{eddy} &\equiv  \kisb{\mathbf{v}^*\cdot\nabla_z p^*} + \kisb{w^*\kirb{\frac{\partial p^*}{\partial z} + \rho^* g}} \label{eq:epsilon_eddy}   \\
&=  \kisb{ \mathbf{U}^*\cdot\kirb{ \nabla p^* + \rho^*\nabla \phi  }  } \;, \label{eq:epsilon_eddy2}
\end{align}
\end{linenomath*}
with an asterisk indicating the deviation from the grid mean. Under hydrostatic balance, the last term in Eq.~\ref{eq:pressure_gradient_work} vanishes.  The last term in Eq.~\ref{eq:epsilon_eddy}, which represents work done against the vertical pressure gradient and cloud-buoyancy forces, arises from parameterized convection, regardless of whether the dynamics are hydrostatic or non-hydrostatic.

The sub-grid momentum equation in anelastic form is \cite<e.g.,>{jeevanjee_effective_2015,tarshish_buoyant_2018}
\begin{linenomath*}
\begin{equation}\label{eq:momentum_subgrid}
\frac{D\mathbf{U}^*}{Dt} = -\frac{1}{\rho_0}\nabla p^* - \frac{\rho^*}{\rho_0}\nabla \phi + \frac{1}{\rho_0}\nabla\cdot \kitens{\tau}^* \;,
\end{equation}
\end{linenomath*}
where $\rho_0=\rho_0(z)$ is a reference density and $\kitens{\tau}^*$ is the sub-grid viscous stress tensor.  By applying $\mathbf{U}^*\cdot$ and using the tensor identity in Eq.~\ref{eq:tens_id}, we obtain
\begin{linenomath*}
\begin{equation}\label{eq:kinetic_subgrid}
\frac{DK^*}{Dt} =  -\frac{1}{\rho_0}\mathbf{U}^*\cdot\kirb{\nabla p^* +\rho^*\nabla\phi} + \frac{1}{\rho_0}\nabla\cdot\kirb{\kitens{\tau}^*\cdot\mathbf{U}^*} - \delta^* \;,
\end{equation}
\end{linenomath*}
where $K^*\equiv (\mathbf{U}^*\cdot \mathbf{U}^*)/2$ is the sub-grid kinetic energy and $\delta^* \equiv \rho_0^{-1}\kitens{\tau}^*:\nabla\mathbf{U}^*$ is the frictional heating due to sub-grid motions.

Most models assume sub-grid kinetic energy is in steady state because no prognostic memory is allocated for sub-grid momentum\footnote{If a model does account for sub-grid momentum prognostically, its contribution must be included in the Eulerian tendency term of the MSE budget, but this treatment is uncommon.}.  Multiplying Eq.~\ref{eq:kinetic_subgrid} by $\rho_0$ and using the anelastic continuity equation $\nabla \cdot \bigl(\rho_0\mathbf{U}^*\bigr)=0$ gives the flux-divergence form in steady state:
\begin{linenomath*}
\begin{equation}
\nabla\cdot \kirb{\rho_0 K^* \mathbf{U}^*} =  -\mathbf{U}^*\cdot\kirb{\nabla p^* +\rho^*\nabla\phi} +\nabla\cdot\kirb{\kitens{\tau}^*\cdot\mathbf{U}^*} - \rho_0\delta^*\; .
\end{equation}
\end{linenomath*}

Finally, taking the grid-box average, using the definition of $\rho\widetilde{\epsilon}_{eddy}$ in Eq.~\ref{eq:epsilon_eddy2}, and noting that, in general, sub-grid momentum does not exchange horizontally between columns (so the horizontal flux-divergence terms vanish) yields
\begin{linenomath*}
\begin{equation}
\rho \widetilde{\epsilon}_{eddy} = -\frac{\partial }{\partial z}\kisb{ \rho_0 K^* w^* - \kirb{\kitens{\tau}\cdot \mathbf{U}^*}_v}-\rho_0\kisb{\delta^*} \;,
\end{equation}
\end{linenomath*}
which can be merged into $\nabla \cdot \mathbf{F}_t$ and $\rho\delta$ in the MSE (or DSE) budget equation.

In ModelE3, $\rho\widetilde{\epsilon}_{eddy}$ is evaluated within the convective parameterization: when buoyancy triggers sub-grid vertical acceleration, the resulting work and frictional dissipation are computed and applied as heating (or cooling) tendencies to the temperature field, thereby modifying MSE (and DSE) in each layer. This same treatment is required in any model that employs a convective parameterization---hydrostatic or nonhydrostatic---so the term must be computed explicitly within the convection scheme and applied to the temperature tendency.

\section{Direct Derivation of Equation~\ref{eq:flux_advective_mse}}\label{sec:derivation_advective}
Here, we derive Eq.~\ref{eq:flux_advective_mse} by applying the Leibniz integral rule to $\nabla_z\cdot\kiab{h\mathbf{v}}$, as follows:
\begin{linenomath*}
\begin{eqnarray}
\nabla_z\cdot\kiab{h\mathbf{v}} &\equiv& \nabla_z\cdot\frac{1}{g}\int_0^{p_s}h\mathbf{v}\ud p \\
&= & \frac{1}{g}h_s\mathbf{v}_s \cdot \nabla_z p_s + \frac{1}{g}\int_0^{p_s}\mathbf{v}\cdot\nabla_p h \ud p + \frac{1}{g}\int_0^{p_s}h\nabla_p\cdot\mathbf{v}\ud p \;. \label{eq:c2}
\end{eqnarray}
\end{linenomath*}
By applying the mass conservation equation in the $p$-coordinate system, $\nabla_p\cdot\mathbf{v}+\partial \omega_p/\partial p = 0$, we obtain
\begin{linenomath*}
\begin{eqnarray}
\text{The last term of Eq.~\ref{eq:c2}} &= & -\frac{1}{g}\int_0^{p_s}h\frac{\partial \omega_p}{\partial p}\ud p\\
&=& -\frac{1}{g}h_s\omega_{p,s} + \frac{1}{g}\int_0^{p_s}\omega_p\frac{\partial h}{\partial p}\ud p \; , \label{eq:c4}
\end{eqnarray}
\end{linenomath*}
where $\omega_p = 0$ at the top of the atmosphere. By definition, we have $\omega_{p,s}\equiv Dp_s/Dt = \partial p_s/\partial t + \mathbf{v}_s\cdot\nabla_z p_s$. Consequently, by applying this to Eq.~\ref{eq:c4}, along with Eq.~\ref{eq:c2}, we derive
\begin{linenomath*}
\begin{equation}
\nabla_z\cdot\kiab{h\mathbf{v}} = -\frac{1}{g}h_s\frac{\partial p_s}{\partial t} + \frac{1}{g}\int_0^{p_s}\mathbf{v}\cdot\nabla_p h \ud p + \frac{1}{g}\int_0^{p_s}\omega_p\frac{\partial h}{\partial p}\ud p \; .
\end{equation}
\end{linenomath*}

\section{Dry Static Energy Budget and Its Accurate Computation}\label{sec:dse}
By incorporating $D\phi/Dt \equiv gw$ into Eq.~\ref{eq:enthalpy}, we derive the dry static energy (DSE) budget equation:
\begin{linenomath*}
\begin{equation}
\frac{Ds}{Dt} = -\frac{1}{\rho}\nabla\cdot\kirb{\mathbf{R}+\mathbf{F}_t} +L_v\mathcal{C}+L_f\mathcal{F} + L_s\mathcal{D} +\epsilon+ \delta\,,
\end{equation}
\end{linenomath*}
where $s\equiv c_pT +\phi$ represents the DSE. Multiplying by $\rho$ and following a similar derivation to that in Eq.~\ref{eq:column_mse}, we obtain:
\begin{linenomath*}
\begin{equation}\label{eq:col_dse_original}
\frac{\partial \kiab{s}}{\partial t} = -\nabla_z\cdot\kiab{s\mathbf{v}} +R + H +\kiab{L_v\mathcal{C}}+\kiab{L_f\mathcal{F}} + \kiab{L_s\mathcal{D}} +\kiab{\epsilon}+ \kiab{\delta}\;,
\end{equation}
\end{linenomath*}
where the surface value of the vertical component of $\mathbf{F}_t$ is set equal to $H$ in ModelE3. 
This equation is often approximated as \cite<e.g.,>[]{yanai_determination_1973, neelin_modeling_1987}:
\begin{linenomath*}
\begin{equation}
\frac{\partial \kiab{s}}{\partial t} \simeq -\nabla_z\cdot\kiab{s\mathbf{v}} +R + H +L_vP_s\;,
\end{equation}
\end{linenomath*}
where $P_s$ denotes the surface precipitation. In this approximation, we neglect $\kiab{\epsilon}$ and $\kiab{\delta}$ and further assume that all condensed water (liquid, frozen, or deposited ice) is immediately precipitated with no storage in cloud condensate, so that the integrated phase-change terms are represented by $L_v P_s$.

Utilizing the relation, $\kiab{\epsilon} = \partial \kiab{R_dT}/\partial t + \kiab{\widetilde{\epsilon}}$, Eq.~\ref{eq:col_dse_original} can be rearranged into:
\begin{linenomath*}
\begin{equation}
\frac{\partial }{\partial t} \kiab{s-R_dT}= -\nabla_z\cdot\kiab{s\mathbf{v}} +R + H +\kiab{L_v\mathcal{C}}+\kiab{L_f\mathcal{F}} + \kiab{L_s\mathcal{D}} +\kiab{\widetilde{\epsilon}}+ \kiab{\delta}\; ,
\end{equation}
\end{linenomath*}
where $s-R_dT = c_vT + \phi$. Furthermore, under the hydrostatic balance, by employing the relation, $\kiab{\phi} = \kiab{R_dT}+ z_sp_s$, the equation can also be expressed as:
\begin{linenomath*}
\begin{equation}\label{eq:column_dse3}
\frac{\partial }{\partial t}\kirb{ \kiab{c_pT} + z_sp_s}= -\nabla_z\cdot\kiab{s\mathbf{v}} +R + H +\kiab{L_v\mathcal{C}}+\kiab{L_f\mathcal{F}} + \kiab{L_s\mathcal{D}} +\kiab{\widetilde{\epsilon}}+ \kiab{\delta}\;.
\end{equation}
\end{linenomath*}

For facilitating comparison with previous studies, it is useful to express the DSE equation in an advective form within the $p$-coordinate system. By expanding the material derivative in the $p$-coordinate and using the relation from Eq.~\ref{eq:change_coord}, we derive the following:
\begin{linenomath*}
\begin{equation}
\kirb{\frac{\partial s}{\partial t} }_p = - \mathbf{v}\cdot\nabla_p s - \omega_p\frac{\partial s}{\partial p} + g\frac{\partial}{\partial p}\kirb{R_v + F_{t,v}} + L_v\mathcal{C}+L_f\mathcal{F} + L_s\mathcal{D} +\epsilon+ \delta\; .
\end{equation}
\end{linenomath*}
Similar to Eq.~\ref{eq:enthalpy_advection_form}, we can merge $\epsilon$ into the Eulerian tendency and horizontal advection as follows:
\begin{linenomath*}
\begin{equation}
\kicb{\frac{\partial \kirb{c_pT} }{\partial t}}_p = - \mathbf{v}\cdot\nabla_p \kirb{c_pT} - \omega_p\frac{\partial s}{\partial p} + g\frac{\partial}{\partial p}\kirb{R_v + F_{t,v}} + L_v\mathcal{C}+L_f\mathcal{F} + L_s\mathcal{D} +  \delta\; .
\end{equation}
\end{linenomath*}
Upon taking the vertical integration, we obtain:
\begin{linenomath*}
\begin{equation}
\kiab{\kicb{\frac{\partial \kirb{c_pT} }{\partial t}}_p} = - \kiab{\mathbf{v}\cdot\nabla_p \kirb{c_pT}} - \kiab{\omega_p\frac{\partial s}{\partial p}} + R + H + \kiab{L_v\mathcal{C}}+\kiab{L_f\mathcal{F}} + \kiab{L_s\mathcal{D}} + \kiab{\delta} \; .
\end{equation}
\end{linenomath*}

Similar to Eq.~\ref{eq:column_mse_label}, we can associate each term in Eq.~\ref{eq:column_dse3} with the corresponding integration schemes as follows:
\begin{linenomath*}
\begin{equation}
\frac{\partial }{\partial t}\kirb{ \kiab{c_pT} + z_sp_s}=  \underbrace{-\nabla_z\cdot\kiab{s\mathbf{v}} +\kiab{\widetilde{\epsilon}}+ \kiab{\delta}}_{\text{Dynamics}} + \underbrace{H +\kiab{L_v\mathcal{C}}+\kiab{L_f\mathcal{F}} + \kiab{L_s\mathcal{D}} + R}_{\text{Turbulence, Cloud, Radiation}} \;.
\end{equation}
\end{linenomath*}
Therefore, the column MSE flux convergence at time level $n$, $-(\nabla_z\cdot\kiab{s\mathbf{v}})^n$, can be computed using the process increment method as follows:
\begin{linenomath*}
\begin{equation}
-(\nabla_z\cdot\kiab{s\mathbf{v}})^n =  \frac{\kirb{\kiab{c_pT} + z_sp_s}^{n+1}_0 - \kirb{\kiab{c_pT} + z_sp_s}^n_3 }{\Delta t} -\kiab{\delta}^n  - \kiab{\widetilde{\epsilon}}^n \;.
\end{equation}
\end{linenomath*}
Finally, the column-integrated vertical DSE advection is calculated as a residual of the equation:
\begin{linenomath*}
\begin{equation}
\kiab{ \omega_p\frac{\partial s}{\partial p}} = \nabla_z\cdot\kiab{s\mathbf{v}}+\frac{1}{g}s_s\frac{\partial p_s}{\partial t} - \kiab{\mathbf{v}\cdot\nabla_p s } \;,
\end{equation}
\end{linenomath*}
where $g^{-1}s_s\partial p_s/\partial t$ and $\kiab{\mathbf{v}\cdot\nabla_p s}$ are computed in the m-coordinate system.

\section*{Open Research Section}
The ModelE3 simulation data used for making all figures in the study are available at \url{https://portal.nccs.nasa.gov/datashare/giss-publish/pub/tropicalclimate/entppe_4kuni_1yr_Q/}.

\acknowledgments
This work was primarily supported by the NASA Modeling, Analysis, and Prediction Program. K. Inoue was additionally supported by NSF Award 2303505. Resources supporting this work were provided by the NASA High-End Computing (HEC) Program through the NASA Center for Climate Simulation (NCCS) at Goddard Space Flight Center. The authors thank the three anonymous reviewers for their insightful comments, which substantially improved the manuscript.

\bibliography{references}

\end{document}